\documentclass[final,5p,times,twocolumn,authoryear]{elsarticle}

\usepackage{amssymb, amsmath}
\usepackage{lipsum}
\usepackage{multirow}
\usepackage{tabularx}
\usepackage{subcaption}
\usepackage{xcolor}
\usepackage{booktabs}    
\usepackage{tikz}
\usetikzlibrary{arrows.meta,positioning,fit,backgrounds}
\usepackage{orcidlink}
\usepackage{placeins}

\newcommand{\phantomwiki}{%
  \href{https://github.com/matc-thaher/PHANTOM/wiki}{\textsc{phantom} wiki}}

\newcommand{\phantomoct}{%
  \href{https://github.com/matc-thaher/PHANTOM_oct}{\textsc{phantom} octave}}

\newcommand{\phantommat}{%
  \href{https://github.com/matc-thaher/PHANTOM}{\textsc{phantom}}}

\journal{Astronomy $\&$ Computing}

\begin{document}

\begin{frontmatter}



\title{PHANTOM: A MATLAB and Octave Toolbox Connecting Linear Field Statistics to Dark Matter Halo Observables}


\author[first]{Mohammad Abu Thaher Chowdhury \orcidlink{0000-0002-7578-6832}}
\affiliation[first]{organization={Rensselaer Polytechnic Institute},Department of Physics, Applied Physics, and Astronomy
            city={Troy},
            postcode={12180}, 
            state={New York},
            country={USA}}

\begin{abstract}
We present \textsc{phantom} (Profile and Halo Analysis for Numerous Theoretical dark Matter Observables), a public \textsc{matlab} toolbox and \textsc{octave} package for calculations that connect the linear density field to dark matter halo observables. The package combines a flexible cosmology module with linear power spectrum, variance, and correlation-function solvers, and a halo module that covers mass functions, linear bias, density profiles, and concentration-mass relations for cold, warm, and fuzzy dark matter scenarios. All core routines are validated against the Python package \textsc{colossus}, \textsc{hmf}, and \textsc{halomod}, yielding sub-percent agreement for shared models across distances, power spectra, variance, correlation functions, halo mass functions, and density profiles. \textsc{phantom} is organised around a cosmology structure that stores background expansion, growth, and linear power-spectrum handles; this object is constructed once and passed through the call graph, so that halo statistics and halo structure calculations remain consistent by design. From this single entry point, users can obtain field statistics (power spectrum, variance, correlation function), halo statistics (mass functions, linear bias), and halo observables (enclosed mass, circular velocity, projected density, and lensing convergence) on arbitrary user-defined grids. The toolbox targets users whose analysis pipelines are written in \textsc{matlab} or \textsc{octave}, where a validated native implementation of these models has been absent. The code is released under the MIT licence at \phantommat.
\end{abstract}




\end{frontmatter}






\section{Introduction}
\label{sec:introduction}
The internal structure and abundance of dark matter (DM) haloes link cosmological initial conditions to a wide range of galaxy and cluster observables. Halo density profiles, mass definitions, and concentration–mass relations enter dynamical modelling of galaxies, gravitational lensing analyses, and the interpretation of large-scale structure surveys. A large body of work has produced analytic or semi-analytic prescriptions for these quantities, calibrated against collisionless and beyond cold dark matter (CDM) simulations, but in practice, they must be implemented in software before they can be used in data analysis or simulation pipelines. The Python toolkit \textsc{colossus}~\citep{Diemer_2018} standardised many of these calculations in a single package, providing a coherent interface to cosmology, linear field statistics, halo mass functions, bias, concentration, and density profiles. In parallel, a substantial fraction of instrument models, survey pipelines, and fitting codes in astronomy continue to rely on \textsc{matlab}, and there is growing interest in exploring standard cold dark matter ($\Lambda$CDM) and alternative dark matter theories such as warm dark matter (WDM) and fuzzy dark matter(FDM)/scalar field dark matter (SFDM) using existing \textsc{matlab}-based workflows. At present, there is no public, validated \textsc{matlab} or \textsc{octave} toolbox that reproduces the \textsc{colossus} suite of models and extends it to WDM/FDM-inspired transfer functions and halo structure. \textsc{phantom} addresses this directly. It provides a single, validated entry point for \textsc{matlab} and \textsc{octave} users to perform halo analysis across CDM, WDM, and FDM scenarios, within the same analysis pipelines used for instrument modelling, survey forecasting, and simulation post-processing. Work that previously required either porting \textsc{colossus} outputs into \textsc{matlab} by hand, or building piecemeal implementations from published fitting functions, can now be carried out natively within a single code base. 

The architecture of \textsc{phantom} is built around a single cosmology structure that propagates through the full call graph; the layered design and data flow are described in detail in Section~\ref{sec:architecture}. Profile-derived observables such as enclosed mass, circular velocity, projected density, and lensing convergence sit at the end of this hierarchy, depending on the profile layer, but not on any higher module. This means linear-field and halo-structure components can be recombined freely to explore different observables or dark matter scenarios.

In this layout, the cosmology and power-spectrum module provides the natural ``main call'': once the cosmology structure is built, the linear power spectrum, variance, correlation function, and growth factor are available as function handles that feed the halo mass-function, bias, and concentration modules. The present paper describes one specific wiring of these layers aimed at halo observables, but the same building blocks can support a wider set of applications, including custom power spectrum models, alternative filters for excursion set constructions, or future halo model calculations. The layered design isolates the physics of each module and allows new fitting functions or dark matter models to be added by extending a dispatcher interface, without changing user-facing scripts. \textsc{phantom} is distributed as a \textsc{matlab} toolbox and a package in \textsc{octave} via the public \textsc{GitHub} repository under the MIT licence; installation instructions, version requirements, and
function-level documentation are maintained on the \phantomwiki.

In what follows, Sections~\ref{sec:architecture} through~\ref{sec:halo_obs} describe the software architecture, cosmology module, halo statistics, and halo observables built around a single cosmology structure and its data flow from linear field statistics to halo structure. Section~\ref{sec:examples} presents worked examples and Section~\ref{sec:conclusion} summarises the current capabilities of \textsc{phantom} and outlines planned extensions, including additional dark matter scenarios, and improved numerical backends.


\section{Package structure}
\label{sec:architecture}

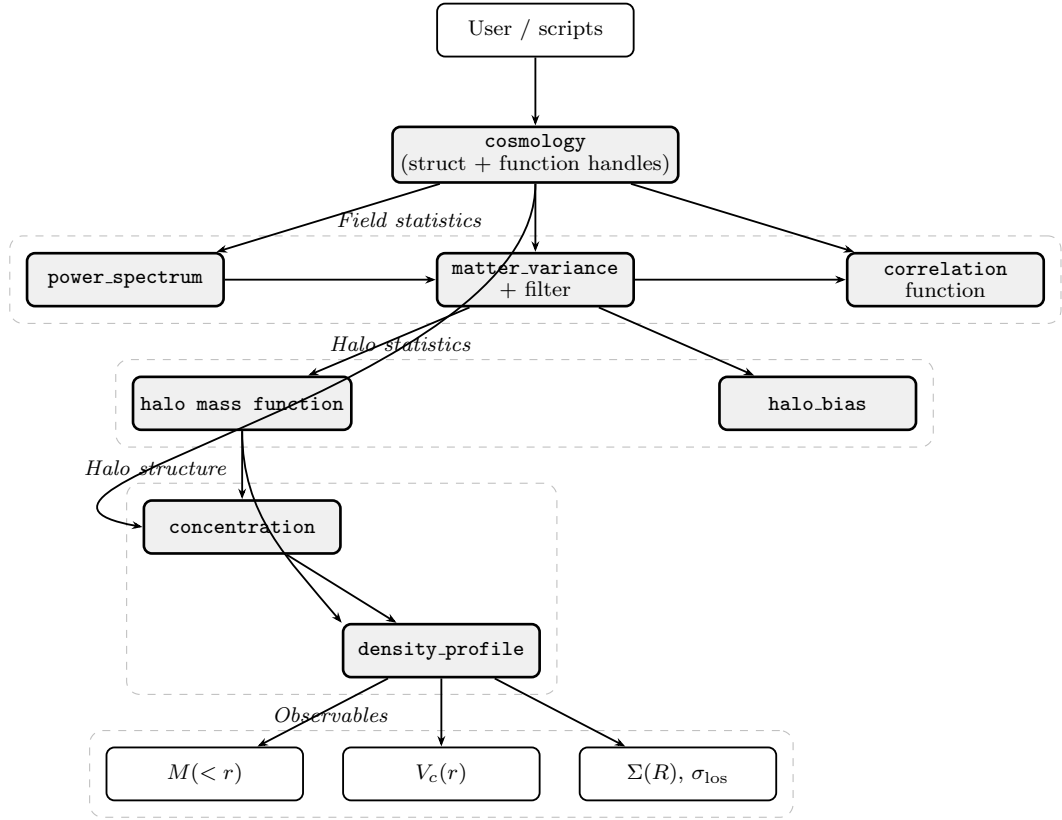
\begin{figure*}
\centering
\begin{tikzpicture}[
  font=\footnotesize,
  node distance=0.55cm and 1.1cm,
  box/.style={draw, rounded corners=3pt, minimum width=2.6cm,
              minimum height=0.7cm, align=center, fill=white,
              line width=0.7pt},
  corebox/.style={box, fill=gray!12, line width=1pt},
  arr/.style={-{Stealth[length=4pt]}, line width=0.7pt}
]

\node[box] (user) {User / scripts};

\node[corebox, below=0.9cm of user]   (cosmo) {\texttt{cosmology}\\(struct + function handles)};

\node[corebox, below left=0.9cm and 2.2cm of cosmo]  (pk)  {\texttt{power\_spectrum}};
\node[corebox, below=0.9cm of cosmo]                  (var) {\texttt{matter\_variance}\\+ filter};
\node[corebox, below right=0.9cm and 2.2cm of cosmo] (xi)  {\texttt{correlation}\\function};

\node[corebox, below left=0.9cm and 1.1cm of var]  (hmf)  {\texttt{halo mass function}};
\node[corebox, below right=0.9cm and 1.1cm of var] (bias) {\texttt{halo\_bias}};

\node[corebox, below=0.9cm of hmf]                          (conc) {\texttt{concentration}};
\node[corebox, below right=0.9cm and 0.0cm of conc]         (prof) {\texttt{density\_profile}};

\node[box, below left=0.9cm and 0.5cm of prof]   (menc)  {$M(<r)$};
\node[box, below=0.9cm of prof]                   (vc)    {$V_c(r)$};
\node[box, below right=0.9cm and 0.5cm of prof]  (sigma) {$\Sigma(R)$, $\sigma_\mathrm{los}$};

\draw[arr] (user) -- (cosmo);

\draw[arr] (cosmo) -- (pk);
\draw[arr] (cosmo) -- (var);
\draw[arr] (cosmo) -- (xi);

\draw[arr] (pk)  -- (var);
\draw[arr] (var) -- (xi);

\draw[arr] (var) -- (hmf);
\draw[arr] (var) -- (bias);

\draw[arr] (hmf)   -- (conc);
\draw[arr] (cosmo.south) to[out=270,in=170] (conc.west);

\draw[arr] (conc)  -- (prof);
\draw[arr] (hmf.south) to[out=270,in=130] (prof.north west);

\draw[arr] (prof) -- (menc);
\draw[arr] (prof) -- (vc);
\draw[arr] (prof) -- (sigma);

\begin{pgfonlayer}{background}
  \node[draw=gray!50, dashed, rounded corners=5pt, inner sep=6pt,
        fit=(pk)(var)(xi), label=above left:{\textit{Field statistics}}] {};
  \node[draw=gray!50, dashed, rounded corners=5pt, inner sep=6pt,
        fit=(hmf)(bias), label=above left:{\textit{Halo statistics}}] {};
  \node[draw=gray!50, dashed, rounded corners=5pt, inner sep=6pt,
        fit=(conc)(prof), label=above left:{\textit{Halo structure}}] {};
  \node[draw=gray!50, dashed, rounded corners=5pt, inner sep=6pt,
        fit=(menc)(vc)(sigma), label=above left:{\textit{Observables}}] {};
\end{pgfonlayer}

\end{tikzpicture}
\caption{\textsc{phantom} software architecture. The cosmology struct
  (shaded boxes) is constructed once and propagated as a function-handle
  container throughout the package. Arrows indicate data flow: the linear
  power spectrum feeds the variance calculation, which in turn drives the
  halo mass function and bias modules. The concentration--mass relation
  takes the halo mass and cosmology as inputs and, together with the
  virial mass, fully specifies the density profile. All profile-derived
  observables — enclosed mass, circular velocity, surface density, and
  line-of-sight velocity dispersion — sit at the end of this hierarchy
  and depend on no module above the profile layer.}
\label{fig:architecture}
\end{figure*}

\textsc{phantom} is organised as a \textsc{matlab} toolbox and \textsc{octave} package, implemented as a flat collection of functions grouped into five logical layers, illustrated in Figure~\ref{fig:architecture}. At the base sits the cosmology struct, which is constructed once via \texttt{cosmology} and passed as an argument to every downstream function; this design avoids global state and allows multiple cosmologies to be run in parallel within the same session. The field-statistics layer computes the linear power spectrum, the matter variance, and the correlation function from the cosmology struct. These outputs feed the halo-statistics layer, which implements the halo mass function and linear halo bias through dispatcher functions that accept a string key and return the requested quantity on any user-supplied grid. The halo-structure layer takes a virial mass and cosmology as inputs, computes the concentration from the chosen fitting formula, and constructs the normalised density profile. All profile-derived observables — enclosed mass, circular velocity, surface density, and line-of-sight velocity dispersion — sit at the end of the call graph and depend on no module above the profile layer. This layered structure means that new models can be added at any level without modifying calling code; contributors need only implement the appropriate dispatcher interface and add an entry to the relevant switch block. 

\subsection{Requirements and availability}
\label{sec:requirements}

\textsc{phantom} is implemented as a flat collection of \textsc{matlab} functions and requires \textsc{matlab} R2021b or later; no additional \textsc{matlab} toolboxes are needed for the analytic transfer-function and halo-model modules. \textsc{octave} support is provided through \phantomoct, a parallel distribution of the core function set compatible with \textsc{octave} version~4.5 and above. Because \phantomoct \ runs on standard \textsc{octave} kernels, it can be executed interactively from a Jupyter notebook either via the \textsc{octave\_kernel} package (v0.38 or later; \citealt{octave_kernel}), a thin wrapper that pipes commands to a system \textsc{octave} process, or via \textsc{xeus-octave} (v0.6.2 or later; \citealt{xeus_octave}), a native Jupyter kernel built on the \texttt{xeus} C++ protocol; both routes let users call \textsc{phantom} functions directly from a notebook cell without leaving a Python-based workflow. The optional \textsc{Python} bridge, used to call \textsc{camb} \citep{Lewis_2000}, requires \textsc{Python}~3.8 or later with the respective packages installed in the active environment. The package is released under the MIT licence at
\phantommat. Once installed, a basic functional check consists of constructing a cosmology structure and evaluating the linear power spectrum on a small wavenumber grid; a reference script for this check is provided in the \texttt{examples/} directory of the repository. The \texttt{tests/} directory contains a suite of tests covering all  \textsc{phantom} functions, including the cosmology module, linear power spectrum, matter variance, correlation function, halo mass function, halo bias, concentration--mass relation, density profiles, and all profile-derived observables; running these tests against a new installation confirms that numerical outputs agree with the validated reference values documented in the repository. All modules include input validation: functions that require model-specific parameters issue a warning and fall back to a documented default when the parameter is absent (for example, the FDM boson mass defaults to $m = 10^{-22}\,\mathrm{eV}$ if \texttt{cosmo.m22} is not set), while functions that depend on external resources raise an error with an explicit message if the required path or pre-computed table is not provided (for example, the \texttt{camb} and \texttt{axioncamb} power-spectrum backends require a valid Python path or a pre-computed \texttt{matter\_power} file, respectively). Full installation instructions and the complete function reference are documented on the \phantomwiki.

\subsection{Computational cost}
\label{subsec:timing}

Table~\ref{tab:timing} reports median wall-clock times for five representative \textsc{phantom} function calls, measured on a single core of an Alienware x15~R1 laptop equipped with an 11th-generation Intel Core i7-11800H (2.30~GHz, 8~cores) and 16~GB of RAM, running MATLAB~R2025a Update~1 on Windows~11 Build~26200. The \textsc{colossus} reference times were obtained on the same machine under Python~3.11, using identical grid sizes and the same \texttt{timeit}-based median-of-five methodology.

\textsc{phantom} and \textsc{colossus} show broadly comparable runtimes across all five modules, with \textsc{phantom} faster than \textsc{colossus} for the power spectrum and within a factor of a few for all other quantities. The remaining differences reflect a fundamental distinction in numerical strategy rather than implementation efficiency. \textsc{colossus} pre-computes interpolation tables at initialisation and evaluates all derived quantities through fast table look-up; \textsc{phantom} performs direct numerical integration for the variance $\sigma(R,z)$ and uses \texttt{fzero}-based root-finding for the concentration--mass relation at every function call. These choices prioritise numerical transparency and self-consistency:the same quadrature scheme and the same collapse-threshold evaluation are applied regardless of how the function is invoked,at a modest cost in per-call overhead relative to an interpolation-based backend.

The practical implication is that both codes are fast enough for typical research workflows. \textsc{colossus} itself demonstrates that interpolation achieves sub-percent accuracy \citep{Diemer_2018}, so the speed gap should not be read as a statement about correctness; it reflects a deliberate design choice to prioritise numerical transparency over per-call overhead. The present runtimes are therefore well within the requirements of the intended use cases of \textsc{phantom}: single-halo analysis, profile fitting, science-demonstration scripts, and moderate-sized parameter sweeps over mass and redshift grids.

\begin{table*}
  \centering
  \caption{Wall-clock times for representative \textsc{phantom} and \textsc{colossus} function calls, evaluated on a single core of an Alienware x15~R1 (Intel Core i7-11800H, 2.30~GHz) running MATLAB~R2025a Update~1 (\textsc{phantom}) and Python~3.11 (\textsc{colossus}) on Windows~11 Build~26200. Grid sizes are representative of typical research use. Times are medians over five repeated calls measured with \texttt{timeit}. See Section~\ref{subsec:timing} for a discussion of the runtime differences.}
  \label{tab:timing}
  \begin{tabular*}{\textwidth}{@{\extracolsep{\fill}} llrrr}
  \hline\hline
  Module & Model & Grid &
    \textsc{phantom} (s) & \textsc{colossus} (s) \\
  \hline
  Power spectrum  & \texttt{eh98}       & $10^{3}$~pts        & 0.0002 & 0.0005 \\
  Variance        & top-hat             & $5\times10^{2}$~pts & 0.0053 & 0.0001 \\
  HMF             & \texttt{tinker08}   & $5\times10^{2}$~pts & 0.0117 & 0.0003 \\
  Concentration   & \texttt{ishiyama21} & $5\times10^{2}$~pts & 0.1140 & 0.0034 \\
  Surface density & NFW                 & $2\times10^{2}$~pts & 0.0013 & 0.0002 \\
  \hline\hline
  \end{tabular*}
\end{table*}

\section{Cosmology}
\label{sec:cosmology}

\subsection{Background cosmology and initialisation}
\label{subsec:cosmo_init}

\begin{table*}
\centering
\footnotesize
\setlength{\tabcolsep}{5pt}
\caption{Pre-set flat $\Lambda$CDM cosmologies implemented in \textsc{phantom}. All models are spatially flat with a cosmological constant, except \texttt{EdS}, which has no dark energy. The parameter $h$ is defined via $H_0 = 100\,h\;\mathrm{km\,s^{-1}\,Mpc^{-1}}$.}
\label{tab:cosmologies}
\resizebox{\textwidth}{!}{%
\begin{tabular}{p{2.8cm} p{0.9cm} p{1.1cm} p{1.1cm} p{1.0cm} p{1.0cm} p{9.0cm}}
\hline\hline
ID & $h$ & $\Omega_{\rm m}$ & $\Omega_{\rm b}$ & $n_{\rm s}$ & $\sigma_8$ & Comment (Reference) \\
\hline
\texttt{UCHUU}         & 0.6774 & 0.3089 & 0.0486 & 0.9667 & 0.8159 & Uchuu simulation \citep{Ishiyama_2021} \\
\texttt{Planck18}      & 0.6766 & 0.3111 & 0.0490 & 0.9665 & 0.8102 & Planck best fit, with BAO \citep{Planck_2020} \\
\texttt{Planck18-only} & 0.6736 & 0.3153 & 0.0493 & 0.9649 & 0.8111 & Planck-only best fit \citep{Planck_2020} \\
\texttt{Planck15}      & 0.6774 & 0.3089 & 0.0486 & 0.9667 & 0.8159 & Planck best fit, with external data \citep{Planck_2016} \\
\texttt{Planck15-only} & 0.6781 & 0.3080 & 0.0484 & 0.9677 & 0.8149 & Planck-only best fit \citep{Planck_2016} \\
\texttt{Planck13}      & 0.6777 & 0.3071 & 0.0483 & 0.9611 & 0.8288 & Planck best fit, with external data \citep{Planck_2014} \\
\texttt{Planck13-only} & 0.6711 & 0.3175 & 0.0490 & 0.9624 & 0.8344 & Planck-only best fit \citep{Planck_2014} \\
\texttt{WMAP9}         & 0.6932 & 0.2865 & 0.0463 & 0.9608 & 0.8200 & Best fit, combined data \citep{Hinshaw_2013} \\
\texttt{WMAP9-only}    & 0.6970 & 0.2814 & 0.0464 & 0.9710 & 0.8200 & Max.\ likelihood, WMAP only \citep{Hinshaw_2013} \\
\texttt{WMAP9-ml}      & 0.6970 & 0.2821 & 0.0461 & 0.9646 & 0.8170 & Max.\ likelihood, combined data \citep{Hinshaw_2013} \\
\texttt{WMAP7}         & 0.7020 & 0.2743 & 0.0458 & 0.9680 & 0.8160 & Best fit, with BAO and $H_0$ \citep{Komatsu_2011} \\
\texttt{WMAP7-only}    & 0.7030 & 0.2711 & 0.0451 & 0.9660 & 0.8090 & Max.\ likelihood, WMAP only \citep{Komatsu_2011} \\
\texttt{WMAP7-ml}      & 0.7040 & 0.2715 & 0.0455 & 0.9670 & 0.8100 & Max.\ likelihood, with BAO and $H_0$ \citep{Komatsu_2011} \\
\texttt{WMAP5}         & 0.7050 & 0.2732 & 0.0456 & 0.9600 & 0.8120 & Best fit, with BAO and SNe \citep{Komatsu_2009} \\
\texttt{WMAP5-only}    & 0.7240 & 0.2495 & 0.0432 & 0.9610 & 0.7870 & Max.\ likelihood, WMAP only \citep{Komatsu_2009} \\
\texttt{WMAP5-ml}      & 0.7020 & 0.2769 & 0.0459 & 0.9620 & 0.8170 & Max.\ likelihood, with BAO and SNe \citep{Komatsu_2009} \\
\texttt{WMAP3}         & 0.7350 & 0.2342 & 0.0413 & 0.9510 & 0.7420 & Best fit, WMAP only \citep{Spergel_2007} \\
\texttt{WMAP3-ml}      & 0.7320 & 0.2370 & 0.0414 & 0.9540 & 0.7560 & Max.\ likelihood, WMAP only \citep{Spergel_2007} \\
\texttt{WMAP1}         & 0.7200 & 0.2700 & 0.0463 & 0.9900 & 0.9000 & Best fit, WMAP only \citep{Spergel_2003} \\
\texttt{WMAP1-ml}      & 0.6800 & 0.3136 & 0.0497 & 0.9700 & 0.9000 & Max.\ likelihood, WMAP only \citep{Spergel_2003} \\
\texttt{illustris}     & 0.7040 & 0.2726 & 0.0456 & 0.9630 & 0.8090 & Illustris simulation \citep{Vogelsberger_2014} \\
\texttt{bolshoi}       & 0.7000 & 0.2700 & 0.0469 & 0.9500 & 0.8200 & Bolshoi simulation \citep{klypin_2011} \\
\texttt{Planck-multidark} & 0.6780 & 0.3070 & 0.0480 & 0.9600 & 0.8290 & MultiDark-Planck simulation \citep{klypin_2016} \\
\texttt{millennium}    & 0.7300 & 0.2500 & 0.0450 & 1.0000 & 0.9000 & Millennium simulation \citep{Springel_2005} \\
\texttt{EdS}           & 0.7000 & 1.0000 & 0.0000 & 1.0000 & 0.8200 & Einstein-de Sitter \citep{Diemer_2018} \\
\hline
\end{tabular}}
\end{table*}

Our implementation follows a standard Friedmann–Lemaître–Robertson–Walker (FLRW) framework with cold dark matter and a cosmological constant, extended to non-flat geometries and alternate dark energy models where needed. The relevant background relations can be found in standard cosmology textbooks \citep[e.g.][]{Dodelson_2003,Mo_2010, Baumann_2022}, and we adopt the same notation conventionally used for numerical cosmology work. The user selects a set of cosmological parameters as listed in Table~\ref{tab:cosmologies}, which close the system once a dark energy model and curvature are specified.

The code stores dimensionless Hubble constant $(h=\frac{H_0}{100\,\,{\rm km\,s^{-1}\,Mpc^{-1}}})$, dimensionless density parameters (matter $(\Omega_{\rm m})$, baryon $(\Omega_{\rm b})$, radiation $(\Omega_{\rm r})$, dark energy $(\Omega_{\Lambda})$, curvature $(\Omega_{\rm k})$), the primordial power spectrum index $n_{\rm s}$, the power spectrum normalization $(\sigma_8)$ and related derived quantities such as the present-day mean matter density \(\rho_{\rm m,0}=\Omega_{\rm m}\rho_{\rm crit,0}\) and cold dark matter dimensionless density parameter $\Omega_c=\Omega_m - \Omega_b$. The respective densities can be calculated using $\rho_i = \rho_{crit}\,\Omega_i$ where $i=m,b,r,k, \Lambda\ etc.$. The critical density and all other background quantities follow directly from the input parameters and the Friedmann equation. For convenience, we provide a list of flat $\Lambda$CDM cosmologies, see Table~\ref{tab:cosmologies}. Users can supply the parameters listed in Table~\ref{tab:cosmologies} manually; all derived quantities --- including $\rho_{{\rm c},0}$, $\Omega_c$, and the dark energy evolution factor $f_{\rm de}(z)$ --- are computed internally from these inputs.

The dimensionless expansion rate
\(
E(z) = H(z)/H_0
\)
is given by
\begin{equation}
\begin{split}
E(z) = \Big[\,\Omega_{\rm m}(1+z)^3 + \Omega_{\Lambda}\,f_{\rm de}(z) +       \,\Omega_{\rm k}(1+z)^2 \\
        + \Omega_{\rm r}(1+z)^4\,\Big]^{1/2},
\end{split}
\label{eq:expnsn_rt}
\end{equation}
where the radiation density $\Omega_{\rm r}$ can be included if requested \citep[]{Dodelson_2003, Baumann_2022}. The radiation component accounts for photons and massless neutrinos:
\begin{equation}
\Omega_{\rm r} = \Omega_{\gamma} + \Omega_{\nu}\,,
\end{equation}
where photon density parameter, $\Omega_{\gamma} = 4.48 \times 10^{-7} T_{\rm CMB}^4 / h^2$ and neutrino density parameter, $\Omega_{\nu} = \frac{7}{8} \left(\frac{4}{11}\right)^{1/3} N_{\rm eff} \Omega_{\gamma}$. By default, $T_{\rm CMB} = 2.7255$ K~\citep{Fixsen_2009, Planck_2016} and $N_{\rm eff} = 3.046$~\citep{Mangano_2002, Salas_2016, Planck_2020}. Radiation is omitted by default and included only when the user requests relativistic species. For flat cosmologies ($\Omega_{\rm k} = 0$), the dark energy density parameter is set to $\Omega_{\Lambda} = 1 - \Omega_{\rm m} - \Omega_{\rm r}$ if not supplied. For non-flat cosmologies, $\Omega_{\Lambda}$ must be specified explicitly and $\Omega_{\rm k} = 1 - \Omega_{\Lambda}- \Omega_{\rm m} - \Omega_{\rm r}$.

The dark energy evolution factor $f_{\rm de}(z) = exp\left[3 \int_0^z \frac{1 + w(z')}{1 + z'} dz'\right]$ depends on the selected model, which gives,
\begin{equation}
f_{\rm de}(z) = 
\begin{cases}
1 & \Lambda\text{CDM}\\
(1+z)^{3(1+w_0)} & w\text{CDM}\\
(1+z)^{3(1+w_0+w_{\rm a})} \exp\left(-\frac{3 w_{\rm a} z}{1+z}\right) & \text{CPL}\,.
\end{cases}
\end{equation}
The Chevallier-Polarski-Linder (CPL) parametrization follows \citet{Chevallier_2001,Linder_2003}, where $w(a) = w_0 + w_{\rm a}(1-a)$. The default is $w_0 = -1$ and $w_{\rm a} = 0$.
The critical density is evaluated as $\rho_c (z) =  \rho_{c,0} E^2(z)$, whereas the other densities $\rho_i$($i = m, b, \Lambda, r, c$), can be computed from their $z=0$ values using function handles in built cosmology and the redshift scalings implied by Eq.~\eqref{eq:expnsn_rt}. All redshift-dependent quantities are evaluated through the expansion rate, which phantom computes by direct numerical integration at each call. Lookback time and age of the universe are computed by direct numerical integration of the background expansion. The lookback time to redshift $z$ is
\begin{equation}
t_{\rm L}(z) = \frac{1}{H_0} \int_0^z \frac{{\rm d}z'}{(1+z') E(z')}\,,
\end{equation}
and the age of the universe at redshift $z$ is $t(z) = t_0 - t_{\rm L}(z)$ where $t_0$ is obtained by integrating to a high redshift $z_{\rm max}$ (default $10^4$). The line-of-sight comoving distance is
\begin{equation}
d_{\rm c}(z) = \frac{c}{H_0} \int_0^z \frac{{\rm d}z'}{E(z')}\,,
\end{equation}
where $c = 299\,792.458$ km s$^{-1}$. This is returned in comoving Mpc/$h$ by multiplying by $h$.

The transverse comoving distance depends on the spatial curvature. For flat cosmologies ($|\Omega_{\rm k}| < 10^{-12}$),
\begin{equation}
d_{\rm M}(z) = 
\begin{cases}
d_{\rm c}(z) & \forall\  \Omega_{\rm k} = 0\\
\frac{D_{\rm H}}{\sqrt{\Omega_{\rm k}}} \sinh\left(\sqrt{\Omega_{\rm k}} \frac{d_{\rm c}}{D_{\rm H}}\right) & \forall\  \Omega_{\rm k} > 0\\
\frac{D_{\rm H}}{\sqrt{|\Omega_{\rm k}|}} \sin\left(\sqrt{|\Omega_{\rm k}|} \frac{d_{\rm c}}{D_{\rm H}}\right) & \forall\  \Omega_{\rm k} < 0\,,
\end{cases}
\end{equation}
where $D_{\rm H} = c / H_0$ in Mpc/$h$. The angular diameter distance $(d_A (z) )$ and luminosity distance $d_L (z)$ follow from the transverse comoving distance:
\begin{equation}
d_{\rm A}(z) = \frac{d_{\rm M}(z)}{1+z}\,, \quad d_{\rm L}(z) = d_{\rm M}(z) (1+z)\,.
\end{equation}

This distance is referred to as the “transverse
comoving distance” (e.g., ~\citep{Hogg_2000}), but a number of other
terms are used in the literature, e.g., “comoving angular
diameter distance”~\citep{Dodelson_2003}, “comoving coordinate
distance”~\citep{Mo_2010}. All distance integrals are evaluated by direct numerical integration at each function call. This approach is less sensitive to interpolation grid choices than interpolation table based methods, though slower. The cosmology structure stores input parameters as scalar fields and derived quantities as function handles, so downstream modules can call any redshift dependent quantity on demand without any pre-computation step.

\subsection{Linear growth factor}
\label{subsec:growth}

The linear growth factor $D(z)$ provides the link between the background expansion and the normalisation of the matter power spectrum $P(k,z)$, entering all variance ($\sigma$) and correlation ($\xi$) calculations through $P(k,z) \propto D^2(z)$ relative to the $z=0$ spectrum. In all cases the growth factor is normalised such that $D(0)=1$.

In a matter-dominated cosmology, linear perturbations obey
\begin{equation}
\frac{d^2D}{da^2}
+ \frac{3}{2a}\left(1 - \frac{1}{3}\frac{d\ln H}{d\ln a}\right)\frac{dD}{da}
- \frac{3}{2}\frac{\Omega_m H_0^2}{a^5 H^2(a)}\,D = 0,
\label{eq:growth_ode}
\end{equation}
which follows from the continuity and Euler equations combined with the Friedmann equation for pressureless matter \citep[e.g.][]{Peebles_1980, Dodelson_2003, Baumann_2022}.

For the standard case of flat $\Lambda$CDM with negligible radiation, \textsc{phantom} uses the computationally inexpensive \citet{Eisenstein_1999} fitting function; the full analytic form is given in Appendix~\ref{app:growth}. When curvature or relativistic components are included, the code integrates the Heath-Peebles integral \citep{Heath_1977,Peebles_1980}
\begin{equation}
D(z) = \frac{5\Omega_m}{2}\,\frac{H(z)}{H_0}
       \int_z^{\infty} \frac{(1+z')}{[H(z')/H_0]^3}\,dz',
\label{eq:heath}
\end{equation}
valid for arbitrary spatial curvature provided dark energy behaves as a cosmological constant. For non-negligible radiation or a time-varying dark energy equation of state, two further solvers are available; their formulations are collected in Appendix~\ref{app:growth}.

In practice, the growth module selects among these four solutions according to the cosmological content of the model: the \citealt{Eisenstein_1999} approximation for standard flat $\Lambda$CDM, the Heath-Peebles integral for non-flat geometries, a radiation-aware hybrid for cases where relativistic species are included, and a Linder-Jenkins ODE for evolving dark energy. The default flat $\Lambda$CDM path is fast and accurate for standard use cases, while the extended solvers cover the remaining parameter space without any change to user-facing scripts.

\subsection{Linear power spectrum}
\label{subsec:powerspec}

The linear matter power spectrum $P(k,z)$ is constructed from three ingredients: a primordial power law, a transfer function, and the linear growth factor of Section~\ref{subsec:growth}. For a CDM cosmology, the $z=0$ spectrum is
\begin{equation}
\label{eq:Pk0_CDM}
P_{\rm CDM}(k) = A\,k^{n_{\rm s}}\,T_{\rm CDM}^2(k)\,,
\end{equation}
where $n_{\rm s}$ is the scalar spectral index and $A$ is a normalization constant fixed by requiring
\begin{equation}
\sigma_8 \equiv \sigma(R=8\,h^{-1}{\rm Mpc},\,z=0)
\end{equation}
to match the input value (Table~\ref{tab:cosmologies}). The redshift dependence is then given entirely by the growth factor,
\begin{equation}
\label{eq:Pkz}
P(k,z) = P_0(k)\,\left[\frac{D(z)}{D(0)}\right]^2.
\end{equation}

Other dark matter models modify the CDM spectrum through a scale-dependent suppression transfer function $T_{\rm s}(k)\in(0,1]$. For both warm dark matter (WDM) and fuzzy dark matter (FDM), the $z=0$ spectrum takes the form
\begin{equation}
\label{eq:Pk_suppressed}
P_0(k) = T_{\rm s}^2(k)\,P_{\rm CDM}(k)\,,
\end{equation}
so the CDM baseline Eq.~\eqref{eq:Pk0_CDM} is preserved and only $T_{\rm s}(k)$ differs between models. In the code, this is implemented by composing the suppression and CDM transfers into a single effective transfer $T_{\rm tot}(k) = T_{\rm s}(k)\,T_{\rm CDM}(k)$, after which the same $\sigma_8$ normalization procedure is applied uniformly across all analytic models. The distinct character of these suppressions is illustrated in panel~\ref{subfig:cosmo_Pk_wdm_fdm} of Fig.~\ref{fig:cosmo_validation}: WDM produces a smooth exponential cutoff that begins at $k\sim1\,h\,{\rm Mpc}^{-1}$ and falls steeply relative to CDM, whereas FDM develops strong quasi-periodic oscillations at $k\gtrsim2\,h\,{\rm Mpc}^{-1}$ driven by quantum pressure, suppressing $P(k)$ by more than twenty decades before $k=100\,h\,{\rm Mpc}^{-1}$.

For CAMB \citep{Lewis_2000}, the $z=0$ power spectrum is computed directly by invoking CAMB through an automated Python bridge built into \textsc{phantom}. The user must have Python and the \textsc{camb} package installed; beyond that, the entire workflow --- specifying the wavenumber range, running the Boltzmann solver, and retrieving the output --- is handled from within \textsc{matlab} without any manual data preparation. The resulting spectrum is then interpolated using a shape-preserving monotone cubic scheme in $\ln k$, and no additional $\sigma_8$ rescaling is applied so that the spectrum remains consistent with the Boltzmann solution. For axionCAMB \citep{Holzek_2015}, the $z=0$ power spectrum is read from a pre-computed table and interpolated with the same scheme; no additional $\sigma_8$ rescaling is applied for the same reason. Table~\ref{tab:phantom_models_ph} summarizes all available options.

Panels~\ref{subfig:cosmo_Pk_comparison} and~\ref{subfig:cosmo_Pk_ratio} of Fig.~\ref{fig:cosmo_validation} compare $P(k)$ against \textsc{colossus} \citep{Diemer_2018} for the zero-baryon (EH-zb) and full-baryon (EH-full) Eisenstein \& Hu transfer functions at $z=0$. The two codes are visually indistinguishable across ten decades in $P(k)$, from $\sim10^{-5}$ to $\sim10^{5}\,({\rm Mpc}/h)^3$, over the full range $k=10^{-4}$--$10^{2}\,h\,{\rm Mpc}^{-1}$. The fractional residual \textsc{phantom}/\textsc{colossus} shown in panel~\ref{subfig:cosmo_Pk_ratio} confirms this quantitatively: EH-zb remains within $\pm7\times10^{-5}$ ($0.007\%$) at all wavenumbers. EH-full shows oscillatory residuals of amplitude up to $\sim2\times10^{-4}$ ($0.02\%$) at the BAO scales $k\sim0.015$--$0.5\,h\,{\rm Mpc}^{-1}$, with a peak deviation of $\sim3\times10^{-4}$ ($0.03\%$) near $k\sim0.4\,h\,{\rm Mpc}^{-1}$; these arise from differences in the $k$-grid sampling of the baryon acoustic oscillations between the two codes and are not a physical discrepancy. Outside the BAO range, both models agree to better than $10^{-5}$.

\begin{table*}
\centering
\footnotesize
\setlength{\tabcolsep}{7pt}
\caption{Implemented linear power-spectrum, halo mass function, and halo-bias models in \textsc{phantom}. Each row lists the dispatcher \texttt{key}, the dark-matter scenario, and the primary reference for that model. These entries control the Gaussian field, excursion-set mapping, and large-scale bias modules; full parameter definitions and worked examples are documented in the \phantomwiki.}
\label{tab:phantom_models_ph}
\small
\resizebox{\textwidth}{!}{%
\begin{tabular}{p{2.4cm} p{1.6cm} p{15.8cm}}
\hline\hline
Model key & Scenario & Notes (Reference) \\
\hline
\multicolumn{3}{l}{\textit{Linear power spectrum}} \\
\texttt{eh98}         & CDM & Zero-baryon broadband fit; no BAO wiggles; fast default baseline \citep{Eisenstein_1999}.\\
\texttt{eh98\_full}   & CDM & Full fit with baryons and BAO wiggles \citep{Eisenstein_1998}.\\
\texttt{sugiyama95}   & CDM & Legacy fit with baryon-dependent shape parameter; retained for comparison \citep{Sugiyama_1995}.\\
\texttt{viel05}       & WDM & Thermal-relic suppression applied to a CDM baseline; \texttt{cosmo.m\_wdm\_keV} \citep{Viel_2005}.\\
\texttt{bode01}       & WDM & Adjustable-parameter suppression; \texttt{cosmo.m\_wdm\_keV} \citep{Bode_2001}.\\
\texttt{schive25}     & FDM & Redshift-independent FDM suppression; \texttt{cosmo.m22} \citep{Hu_2000,Schive_2026}.\\
\texttt{camb}         & CDM & $P_0(k)$ via built-in Python--CAMB bridge; requires Python and \textsc{camb} \citep{Lewis_2000}.\\
\texttt{axioncamb}    & FDM & Reads precomputed \texttt{matterpower.dat} from axionCAMB; no renormalization \citep{Holzek_2015,Grin_2022}.\\
\hline
\noalign{\smallskip}
\multicolumn{3}{l}{\textit{Halo mass function}} \\
\texttt{ps}             & CDM & Excursion-set spherical-collapse derivation; no free parameters \citep{Press_1974}.\\
\texttt{st}             & CDM & Ellipsoidal-collapse correction to Press--Schechter; aliases \texttt{sheth99}, \texttt{sheth01} \citep{Sheth_1999}.\\
\texttt{reed03}         & CDM & Early small-scale correction to Press--Schechter; retained for comparison with legacy work \citep{Reed_2003}.\\
\texttt{reed07}         & CDM & Improved CDM fit with non-Gaussian corrections to the excursion-set collapse barrier \citep{Reed_2007}.\\
\texttt{tinker08}       & CDM & $N$-body calibration for SO masses; $\Delta$ relative to mean density; supported range $\Delta = 200$--$3200$ \citep{Tinker_2008}.\\
\texttt{crocce10}       & CDM & FOF fit; power-law redshift dependence of each shape parameter \citep{Crocce_2010}.\\
\texttt{bhattacharya11} & CDM & Redshift-dependent ST-type fit; $M_{200\mathrm{c}}$ and $M_\mathrm{vir}$, calibrated to WMAP7 cosmology \citep{Bhattacharya_2011}.\\
\texttt{courtin11}      & CDM & Virial-overdensity calibration from the Horizon-$4\pi$ simulation \citep{Courtin_2011}.\\
\texttt{angulo12}       & CDM & Recalibrated Sheth--Tormen form fitted to the Millennium-XXL simulation \citep{Angulo_2012}.\\
\texttt{watson13}       & CDM & Separate FOF and SO calibrations; explicit redshift evolution of all shape parameters \citep{Watson_2013}.\\
\texttt{despali16}      & CDM & SO fit using virial-to-critical mass conversion; requires $\Delta_\mathrm{c}$, $\Delta_\mathrm{vir,c}$, $\delta_\mathrm{c}$ \citep{Despali_2016}.\\
\texttt{bocquet16}      & CDM & Hydrodynamical-simulation calibration capturing baryonic suppression of the high-mass end \citep{Bocquet_2016}.\\
\texttt{rodriguezpuebla16} & CDM & Separate parametrisations for distinct haloes and subhaloes \citep{Rodriguez_2016}.\\
\texttt{comparat17}     & CDM & MultiDark simulation calibration; FOF mass definition \citep{Comparat_2017}.\\
\texttt{diemer20}       & CDM & Peak-height and large-scale-environment dependent fit \citep{Diemer_2020}.\\
\texttt{seppi20}        & CDM & Full-sample SO fit from the MultiDark-Planck2 simulation \citep{Seppi_2021}.\\
\texttt{seppi20m}       & CDM & Mass-binned variant of \texttt{seppi20} \citep{Seppi_2021}.\\
\texttt{yung24}         & CDM & Predecessor redshift-dependent calibration; see also \texttt{yung25} \citep{Youg_2024}.\\
\texttt{yung25}         & CDM & Updated redshift-dependent fit; extra arg: $z$ \citep{Yung_2025}.\\
\texttt{fernandezgarcia26} & CDM & Extra args: $M$, $z$, \texttt{mdef}, \texttt{cosmo} \citep{Fernando_2026}.\\
\texttt{fiorilli26}     & CDM & Evolution-mapping model; non-universality encoded via formation-history integral $\tilde{x}$ and local power-spectrum slope $n_\mathrm{eff}$; SO definitions $150\mathrm{m}$--$1600\mathrm{m}$; optional unbound-particle parameter set \citep{Fiorilli_2025}.\\
\texttt{schneider12}    & WDM & CDM baseline suppressed by a power-law factor below the half-mode mass $M_{1/2}$; requires \texttt{cosmo.m\_wdm\_keV} \citep{Schneider_2012}.\\
\texttt{lovell14}       & WDM & Ratio fit $n_\mathrm{WDM}/n_\mathrm{CDM}$ calibrated to WDM $N$-body runs; smooth suppression below the free-streaming mass \citep{Lovel_2014}.\\
\texttt{schive16}       & FDM & CDM baseline with two-parameter suppression $\mathcal{F}(M/M_0,\,\beta_1,\beta_2)$ calibrated to FDM $N$-body simulations; requires \texttt{cosmo.m22} \citep{Schive_2016}.\\
\texttt{du17}           & FDM & Mass-dependent FDM collapse barrier via Sheth--Tormen with $\mathcal{G}(M)\,\delta_\mathrm{c}$; $\delta_\mathrm{c}$ defaults to $1.686$ if not supplied \citep{Du_2016}.\\
\hline
\multicolumn{3}{l}{\textit{Halo bias}} \\[2pt]
\texttt{cole89}        & CDM & Press--Schechter / Cole--Kaiser linear bias $b(\nu)=1+(\nu^2-1)/\delta_{\mathrm{c}}$; constant spherical-collapse barrier \citep{Cole_1989}. \\
\texttt{st}            & CDM & Sheth--Tormen constant-barrier bias calibrated to ellipsoidal collapse; pairs naturally with the ST multiplicity function \citep{Sheth_1999}. \\
\texttt{smt01}         & CDM & Sheth--Mo--Tormen moving-barrier bias model; mass-dependent barrier height tuned to $N$-body simulations \citep{Sheth_2001}. \\
\texttt{jing98}        & CDM & Empirical bias fit including dependence on effective spectral index $n_\mathrm{eff}$; optional \texttt{cosmo} struct for $P(k)$-derived $n_\mathrm{eff}$ \citep{Jing_1998}. \\
\texttt{seljak04}      & CDM & Bias calibration for friends-of-friends haloes; optional \texttt{cosmo} for $(\Omega_\mathrm{m},n_\mathrm{s},\sigma_8,h)$ dependence \citep{Seljak_2004}. \\
\texttt{tinker10}      & CDM & Redshift- and overdensity-dependent SO bias fit; extra args: $\Delta$ (default 200), $z$, \texttt{cosmo}; default CDM choice in \textsc{phantom} \citep{Tinker_2010}. \\
\texttt{bhattacharya11}& CDM & Peak-height bias fit with explicit redshift evolution; extra arg: $z$ (default 0) \citep{Bhattacharya_2011}. \\
\texttt{comparat17}    & CDM & Bias calibration from the MultiDark simulations; covers a wide mass and redshift range for FOF haloes \citep{Comparat_2017}. \\
\texttt{pillepich10}   & CDM & Gaussian-mode peak-background split bias from Pillepich et al.; extra arg: \texttt{cosmo}; non-Gaussian extension available via direct call \citep{Pillepich_2010}. \\
\texttt{wdm}           & WDM & Implicit WDM bias: \citet{Sheth_2001} evaluated on WDM $\sigma(M,z)$; extra args: $M$, $M_{1/2}$ to mask spurious low-mass haloes \citep{Schneider_2012}. \\
\texttt{fdm}           & FDM & Moving-barrier bias \citep{Sheth_2001} evaluated on FDM $\sigma(M,z)$ with $\delta_c^{\rm fdm}(M,z)$; self-consistent with \texttt{du17}; runtime warning near $M_{\rm J}$.\\

\hline
\hline
\end{tabular}}
\end{table*}

\subsection{Variance of the density field}
\label{sec:variance}

Given the linear power spectrum, the variance of the density field smoothed on a comoving scale $R$ is
\begin{equation}
\sigma^2(R,z)
= \frac{1}{2\pi^2}
  \int_0^{\infty} dk\, k^2\, P(k,z)\, W^2(k,R),
\label{eq:variance}
\end{equation}
where $W(k,R)$ is the Fourier transform of a spherically symmetric window function. The integral is evaluated over $\ln k$ using adaptive quadrature, with the growth factor $D(z)$ factored out as in Eq.~\ref{eq:Pkz}. The filter choice is exposed through \texttt{cosmo.varianceFilter} and defaults to a real-space top-hat if not specified.

The real-space top-hat has the Fourier-space window
\begin{equation}
W_{\rm TH}(k,R) =
  \frac{3[\sin(kR) - kR\cos(kR)]}{(kR)^3},
\label{eq:tophat}
\end{equation}
and maps directly to a mass scale via $M = \frac{4}{3}\pi\bar{\rho}_{m,0}R^3$, making it the natural choice for CDM halo statistics. Four additional filters are available for models with scale-dependent power suppression: a Gaussian, a sharp-$k$, a smooth-$k$ \citep{Leo_2018}, and a variable-slope smooth-$k$ \citep{Rocamora_2026}; their functional forms and mass-calibration conventions are given in Appendix~\ref{app:variance}. We recommend the top-hat for CDM, smooth-$k$ for WDM, and vsmk for models with scale-dependent damping such as FDM.

\begin{figure}
    \centering
    \includegraphics[width=\columnwidth]{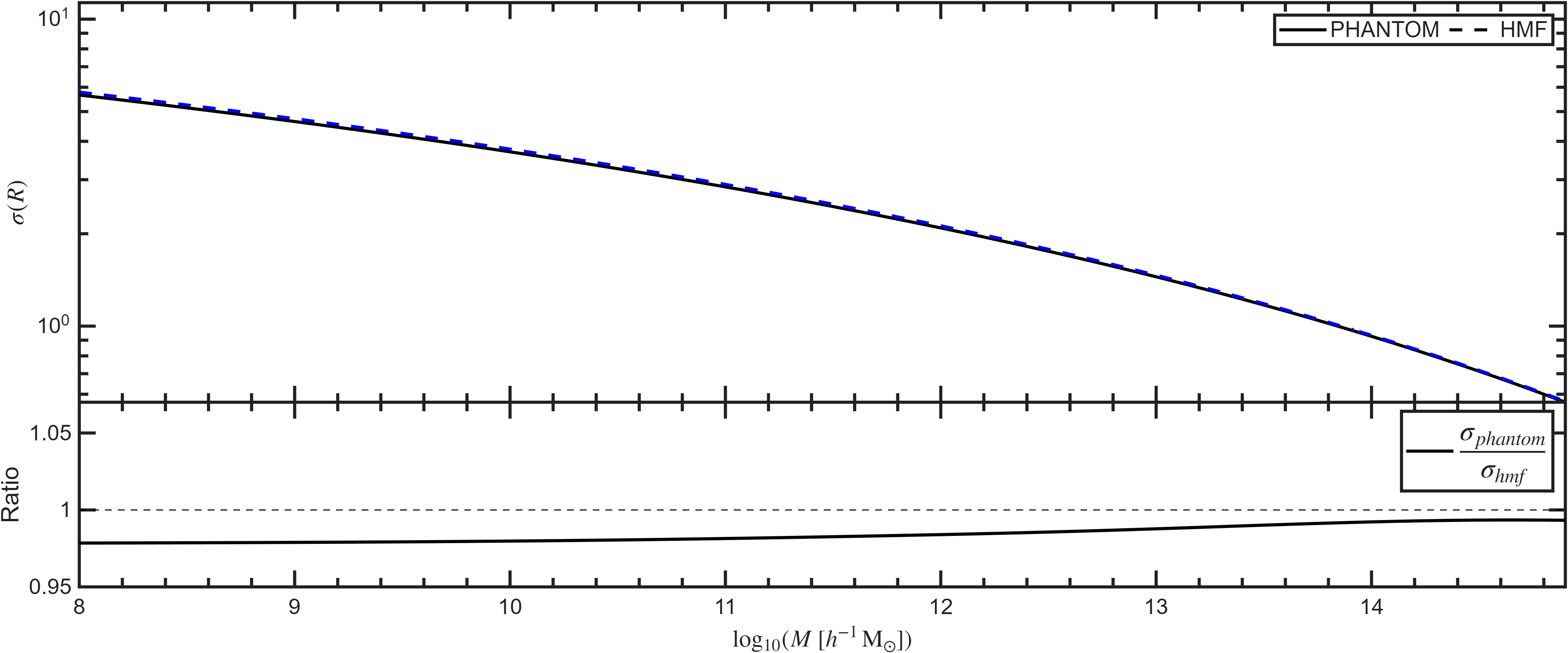}
    \caption{Comparison of the matter variance $\sigma(R)$ between \textsc{phantom} (solid black) and \textsc{hmf} \citep{Murray_2013} (blue dashed) for the EH98 zero-baryon transfer function at $z=0$. The lower sub-panel shows the ratio $\sigma_\textsc{phantom}/\sigma_\textsc{hmf}$; the two codes agree within $2\%$ across the plotted mass range.}
    \label{fig:sigma_hmf}
\end{figure}

The variance $\sigma(R)$ for the \textsc{Planck15} cosmology at $z=0$, computed with the EH-zb and EH-full transfer functions in both \textsc{phantom} and \textsc{colossus} \citep{Diemer_2018}, is shown in Panel~\ref{subfig:cosmo_sigma_comparison}. The four curves lie essentially on top of each other over $R\simeq10^{-2}$--$10^{2}\,h^{-1}{\rm Mpc}$, with the full-baryon variant yielding slightly lower variance on small scales than the zero-baryon case, as expected from the suppression in $P(k)$ at high $k$. The fractional residual ($\sigma_{\rm \textsc{phantom}}(R)/\sigma_{\rm \textsc{colossus}}(R)$) remaining within $0.2\%$ (panel~\ref{subfig:cosmo_sigma_ratio}) across most of the range., with mild oscillations near $R\sim10\,h^{-1}{\rm Mpc}$ that trace the BAO features in the underlying power spectrum, and a sharper excursion at large $R$ where $\sigma(R)$ approaches unity and numerical cancellations in the integrand become more pronounced. As a further cross-check, Fig.~\ref{fig:sigma_hmf} compares $\sigma(R)$ computed with  the EH98 zero-baryon transfer function against the \textsc{hmf} Python package \citep{Murray_2013}; the ratio $\sigma_\textsc{phantom}/\sigma_\textsc{hmf}$ remains within $2\%$ across $10^{8} \lesssim M/(h^{-1}\,\mathrm{M}_\odot) \lesssim 10^{14}$,  with the residual tracing a known difference in the numerical quadrature schemes of the two codes rather than a physical discrepancy. The impact of the window function at fixed power spectrum is examined in Panel~\ref{subfig:cosmo_sigma_filters}, where top-hat, Gaussian, sharp-$k$, smooth-$k$, and vsmk filters are compared for the EH-full model. The top-hat window yields the largest $\sigma(R)$ at small $R$, while the Gaussian and sharp-$k$ filters give progressively lower variance; the smooth-$k$ and vsmk choices follow each other closely with a slightly stronger small-scale suppression, making them well suited to other dark matter model transfer functions with scale-dependent damping such as WDM or FDM.

\subsection{Correlation function}
\label{subsec:correlation}

The linear matter--matter correlation function is the isotropic Fourier transform of the power spectrum,
\begin{equation}
\label{eq:xi}
\xi_{\rm mm}(R,z) = \frac{1}{2\pi^2}\int_0^{\infty}
\mathrm{d}k\,k^2\,P(k,z)\,\frac{\sin(kR)}{kR}.
\end{equation}
This integral converges slowly because the sinc term oscillates rapidly and decays only algebraically at large $kR$, a numerical challenge shared with \textsc{colossus} \citep{Diemer_2018}. The power spectrum enters through the same $P(k,z) = P_0(k)[D(z)/D(0)]^2$ of Section~\ref{subsec:powerspec}, so any choice of transfer model (CDM, WDM, FDM, CAMB, or axionCAMB) propagates automatically into the correlation function without additional modification.

Two numerical methods are available for evaluating Eq.~\eqref{eq:xi}. The first is a direct integration in $\ln k$ using a pre-computed logarithmic $k$-grid and trapezoidal quadrature. Despite its simplicity, this approach evaluates all separations $R$ simultaneously through a vectorized outer product in $k\times R$ space, making it acceptably fast for moderate-sized grids while remaining fully transparent for validation purposes.

The second method uses an FFTLog algorithm \citep{Hamilton_2000}. The key observation is that Eq.~\eqref{eq:xi} is a spherical Bessel transform of order $\mu=1/2$,
\begin{equation}
j_0(kR) = \frac{\sin(kR)}{kR} = \sqrt{\frac{\pi}{2kR}}\,J_{1/2}(kR),
\end{equation}
so that the transform reduces to a standard Hankel transform of order $\mu=1/2$ \citep{Hamilton_2000}. In practice, we apply the discrete logarithmic Hankel transform to the sequence $A(k) = k^{3/2}\,P(k,z)$ and convert the output to $\xi(R)$ via
\begin{equation}
\xi(R,z) = \frac{B(R)}{2\pi^2\,R^{3/2}},
\end{equation}
where $B(R)$ is the transformed sequence evaluated on the conjugate logarithmic grid \citep{Hamilton_2000}. To suppress the ringing introduced by the periodic extension on the logarithmic grid, we choose the central product $k_0 r_0$ to satisfy the low-ringing condition described by \citet{Hamilton_2000}, analogous to the \texttt{krgood} adjustment in \texttt{pyfftlog} \citep{dieter_2024}. The FFTLog method is substantially faster than direct integration when $\xi(R)$ is needed on a dense $R$-grid, at the cost of algorithmic complexity; we therefore recommend validating it against the direct integral method for each new cosmology or power-spectrum model.

The direct-integral method is the default. Users may switch to FFTLog for performance, in which case the integration limits and $k$-grid spacing remain identical to those used in the direct method so that cross-validation is straightforward. The integration range for analytic transfer models spans $k\in[10^{-6},10^{4}]\,h\,{\rm Mpc}^{-1}$; for CAMB and axionCAMB backends, it is restricted to the tabulated $k$-range to avoid extrapolation.

\begin{figure*}
  \centering
  \subfloat[\label{subfig:cosmo_Pk_comparison}]{
    \includegraphics[width=0.32\textwidth]{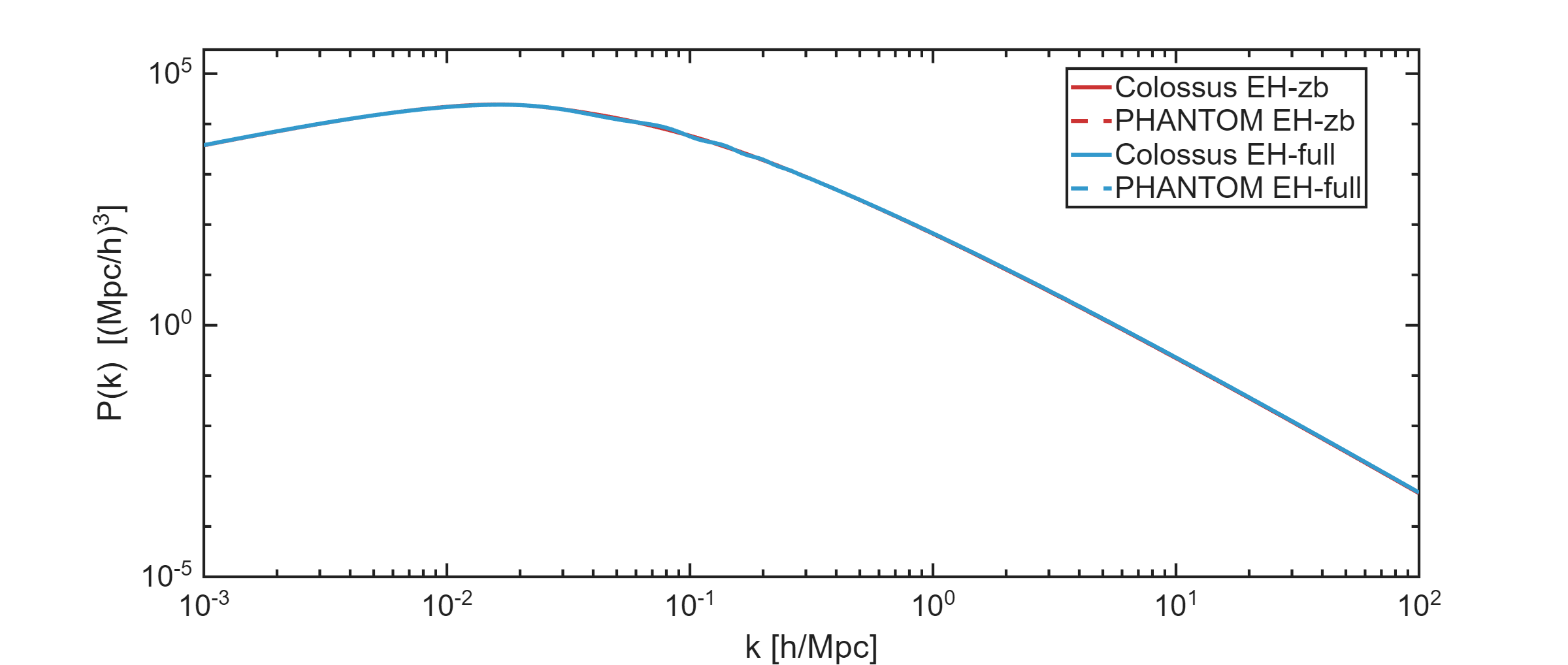}
  }\hfill
  \subfloat[\label{subfig:cosmo_Pk_ratio}]{
    \includegraphics[width=0.32\textwidth]{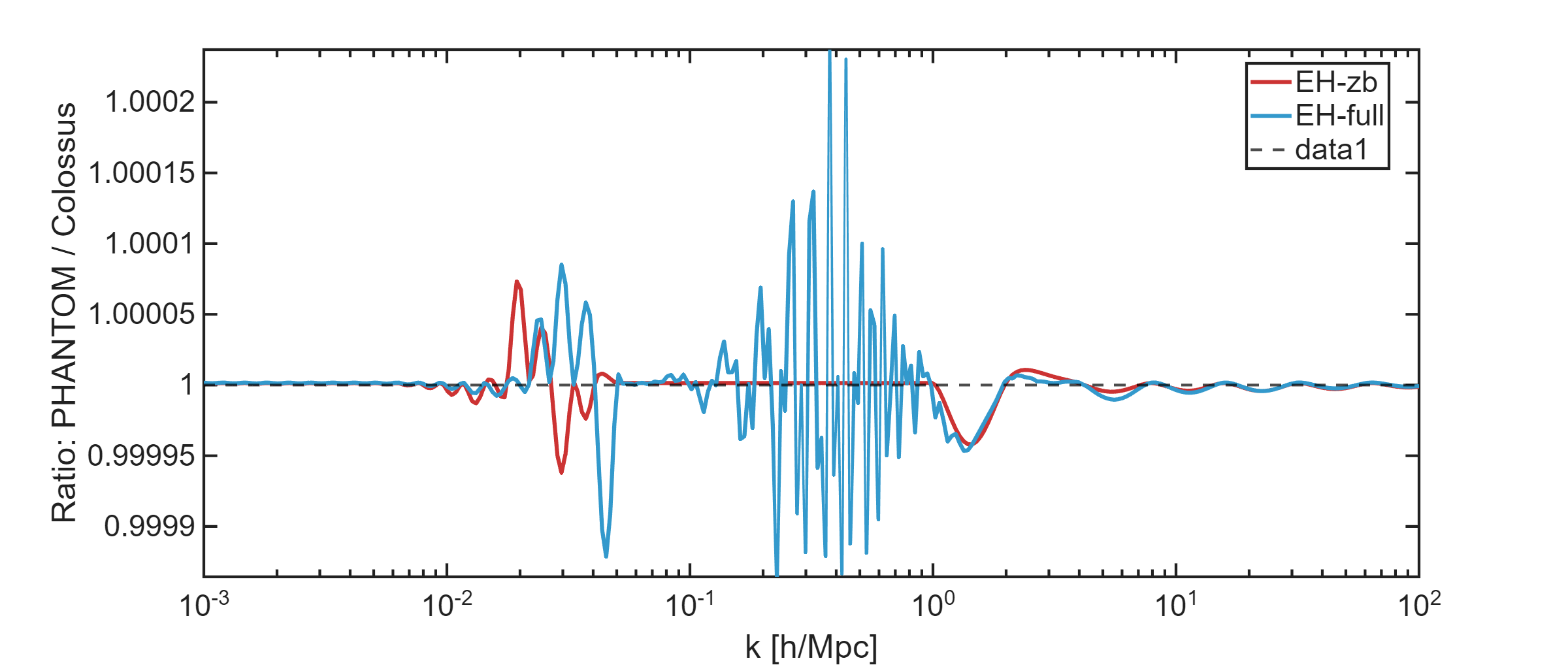}
  }\hfill
  \subfloat[\label{subfig:cosmo_Pk_wdm_fdm}]{
    \includegraphics[width=0.32\textwidth]{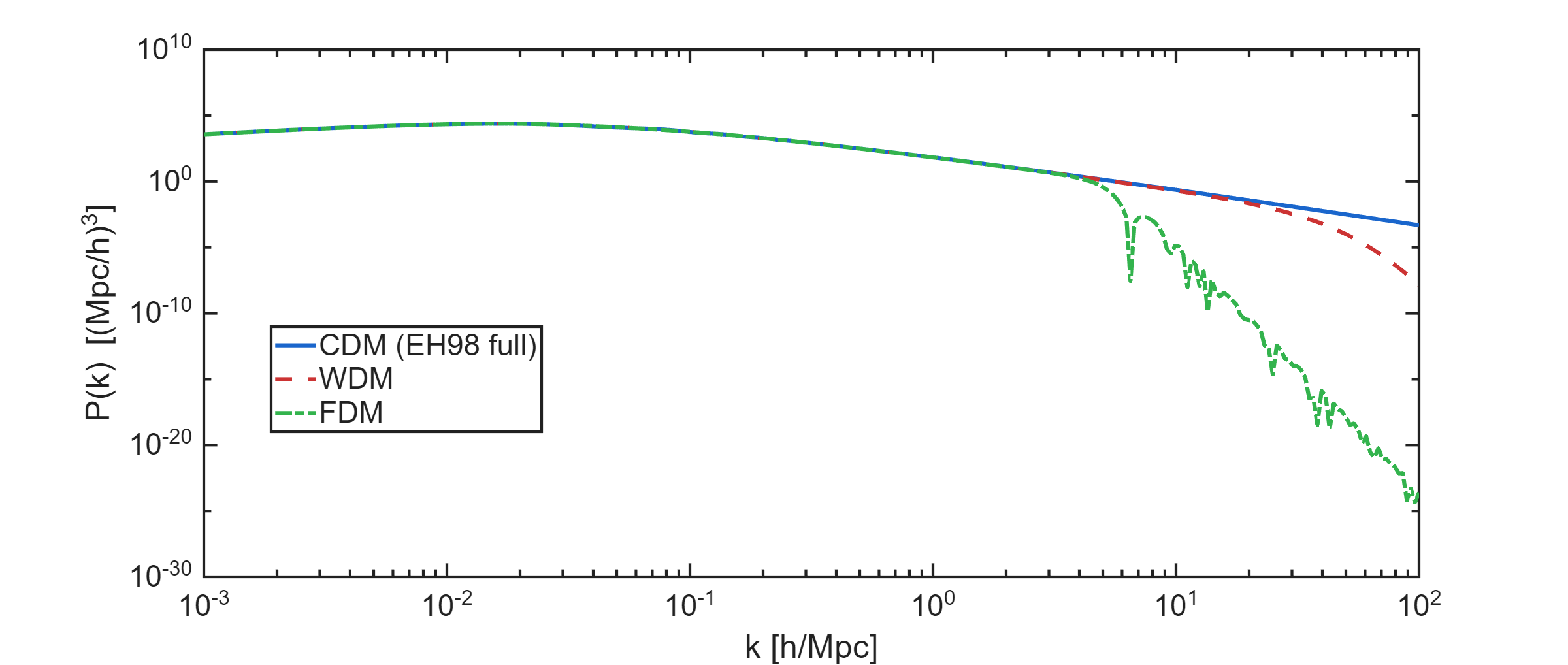}
  }

  \vspace{0.5em}

  \subfloat[\label{subfig:cosmo_sigma_comparison}]{
    \includegraphics[width=0.32\textwidth]{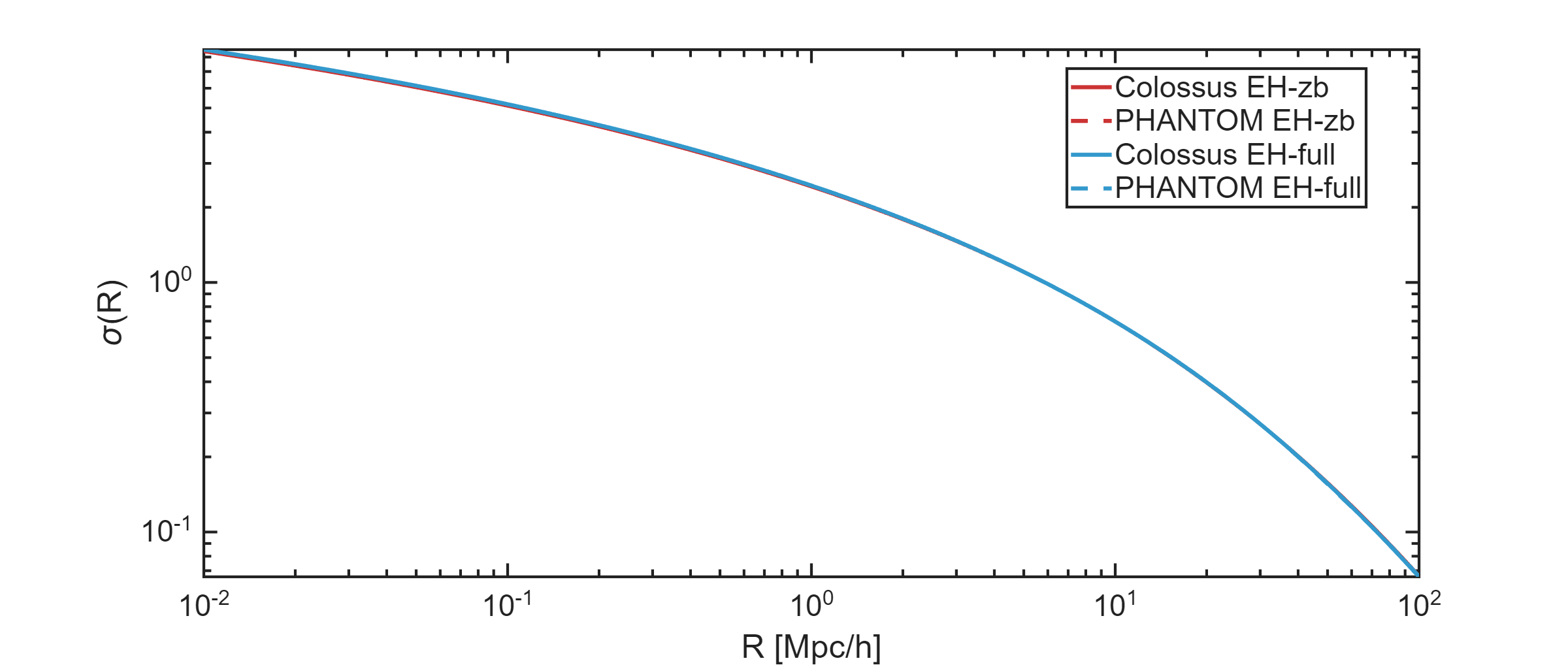}
  }\hfill
  \subfloat[\label{subfig:cosmo_sigma_ratio}]{
    \includegraphics[width=0.32\textwidth]{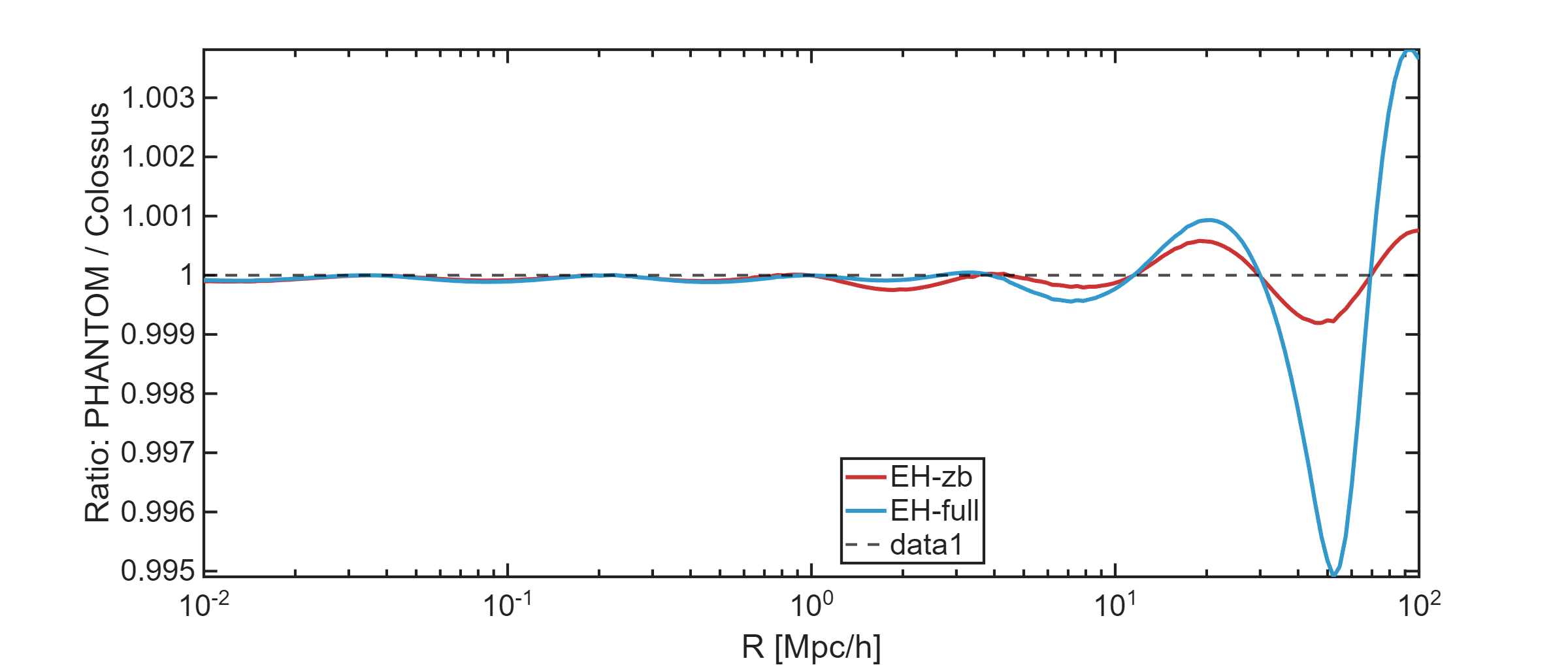}
  }\hfill
  \subfloat[\label{subfig:cosmo_sigma_filters}]{
    \includegraphics[width=0.32\textwidth]{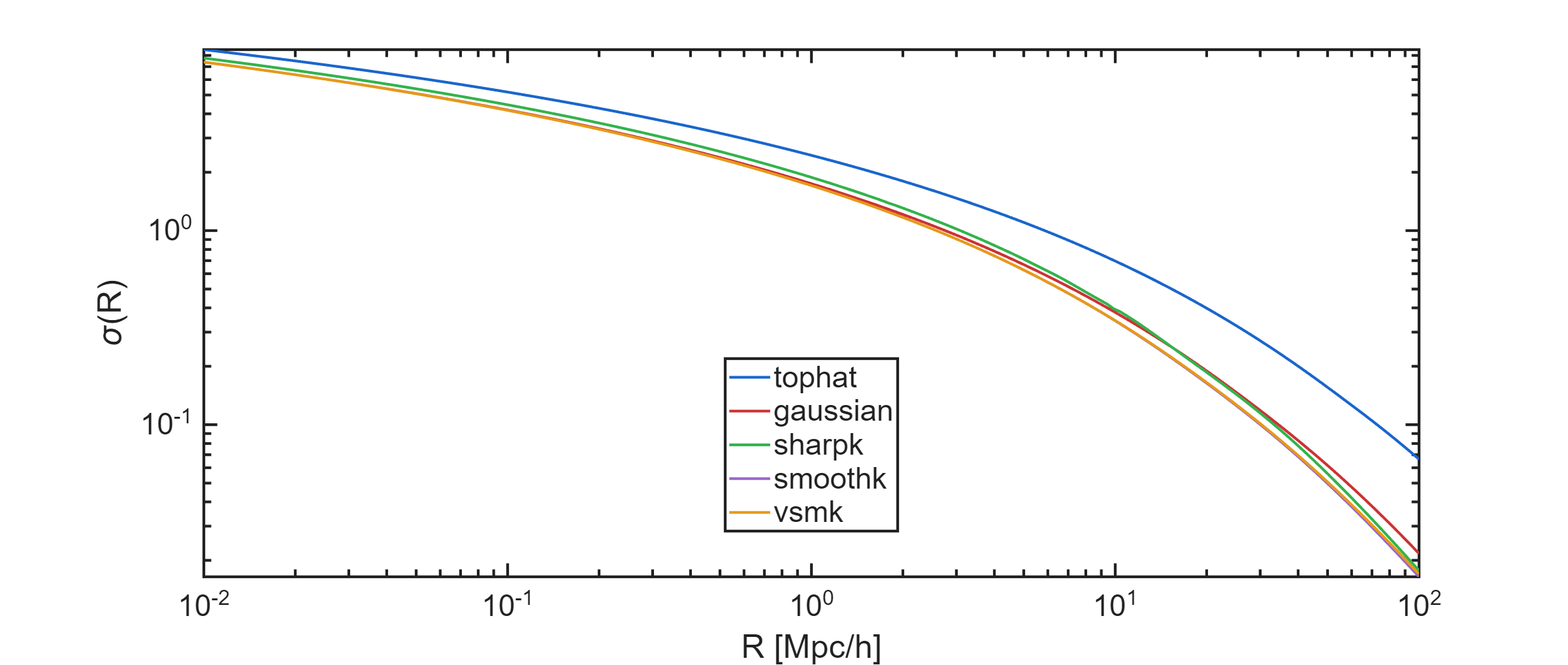}
  }

  \vspace{0.5em}

  \subfloat[\label{subfig:cosmo_xi_comparison}]{
    \includegraphics[width=0.32\textwidth]{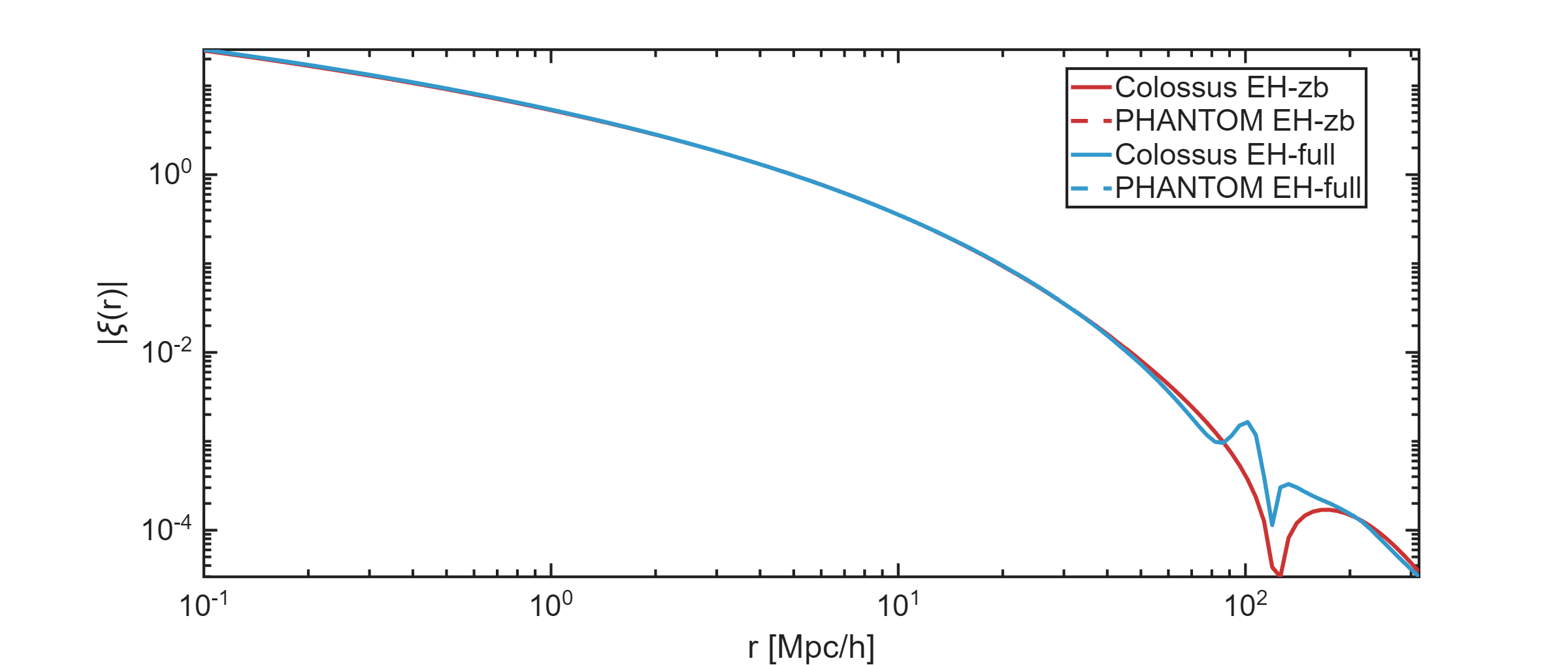}
  }\hfill
  \subfloat[\label{subfig:cosmo_xi_ratio}]{
    \includegraphics[width=0.32\textwidth]{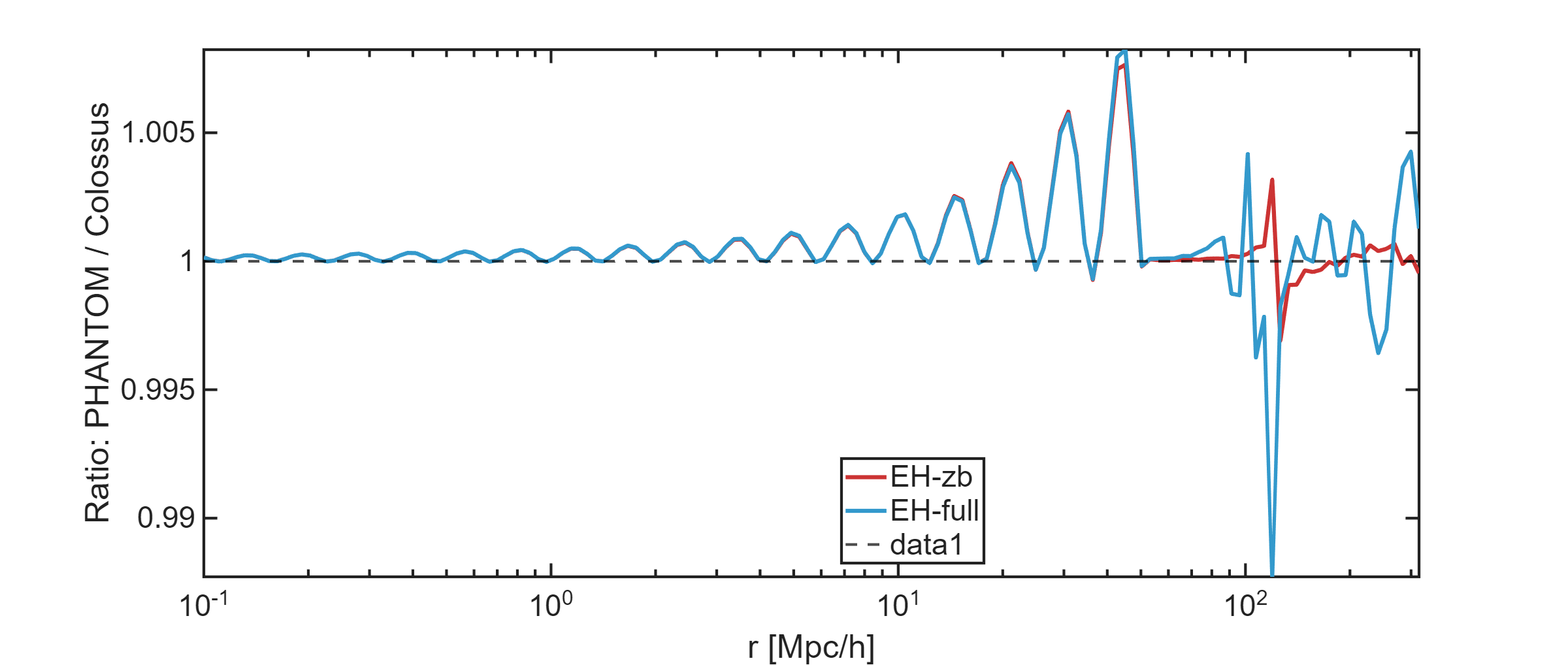}
  }\hfill
  \subfloat[\label{subfig:cosmo_xi_methods}]{
    \includegraphics[width=0.32\textwidth]{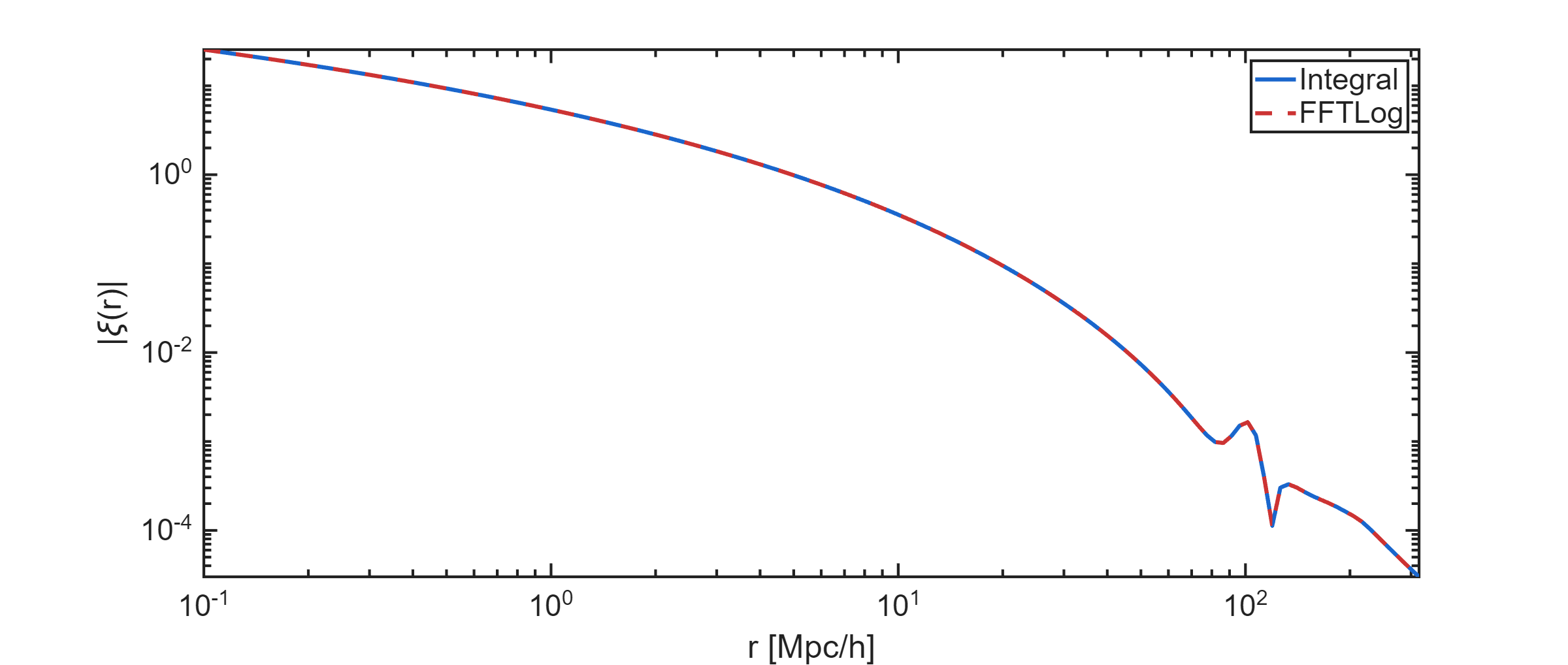}
  }

\caption{
Linear power spectrum, variance, and correlation function from \textsc{phantom} for the \textsc{Planck15} cosmology at $z=0$, compared against \textsc{colossus} \citep{Diemer_2018}, and illustration of additional dark matter models and numerical options.
Top row: \textit{(\ref{subfig:cosmo_Pk_comparison})} linear matter power spectrum $P(k)$ for the zero‑baryon (EH-zb, red) and full‑baryon (EH-full, blue) Eisenstein \& Hu transfer functions; solid lines are \textsc{colossus} and dashed lines are \textsc{phantom}. \textit{(\ref{subfig:cosmo_Pk_ratio})} fractional residual \textsc{phantom}/\textsc{colossus} for $P(k)$; the dashed line marks unity and the ratio remains within $\pm3\times10^{-4}$, with the largest excursions at the BAO scales. \textit{(\ref{subfig:cosmo_Pk_wdm_fdm})} CDM (EH98‑full, blue), WDM (red), and FDM (green) power spectra at $z=0$; WDM shows a smooth cutoff beginning near $k\sim1\,h\,{\rm Mpc}^{-1}$, whereas FDM exhibits strong oscillations and a much sharper small‑scale suppression due to quantum pressure.
Middle row: \textit{(\ref{subfig:cosmo_sigma_comparison})} variance $\sigma(R)$ compared against \textsc{colossus} for EH-zb and EH-full, showing agreement at the sub‑percent level over three decades in $R$. \textit{(\ref{subfig:cosmo_sigma_ratio})} fractional residual \textsc{phantom}/\textsc{colossus} for $\sigma(R)$, demonstrating that both transfer variants remain within $\simeq1\%$ across the plotted range. \textit{(\ref{subfig:cosmo_sigma_filters})} dependence of $\sigma(R)$ on the smoothing window for the EH-full model: real‑space top‑hat (blue), Gaussian (red), sharp‑$k$ (green), smooth‑$k$ (purple), and vsmk (orange), illustrating how sharp‑$k$–type filters reduce small‑scale variance relative to the top‑hat and how the smooth‑$k$ and vsmk filters provide a more gradual transition for suppressed models such as WDM.
Bottom row: \textit{(\ref{subfig:cosmo_xi_comparison})} absolute correlation function $|\xi_{\rm mm}(r)|$ from \textsc{phantom} (dashed) and \textsc{colossus} (solid) for EH-zb and EH-full; the curves overlap over more than four decades in amplitude from $r=0.1$ to $\sim150\,h^{-1}{\rm Mpc}$. \textit{(\ref{subfig:cosmo_xi_ratio})} fractional residual \textsc{phantom}/\textsc{colossus} for $\xi_{\rm mm}$, restricted to $r\lesssim100\,h^{-1}{\rm Mpc}$ where $|\xi|$ is non‑negligible; the spike near $r\sim100\,h^{-1}{\rm Mpc}$ coincides with the zero‑crossing of $\xi$, where the relative error is ill‑defined. \textit{(\ref{subfig:cosmo_xi_methods})} internal comparison of the two \textsc{phantom} methods for $|\xi_{\rm mm}(r)|$: direct trapezoidal integral (blue solid) and FFTLog (red dashed), which are numerically indistinguishable over the range plotted.
}
  \label{fig:cosmo_validation}
\end{figure*}

Figure~\ref{fig:cosmo_validation}e compares $|\xi_{\rm mm}(r)|$ at $z=0$ from \textsc{phantom} and \textsc{colossus} for both EH-zb and EH-full, showing agreement over more than four decades in amplitude from $r=0.1$ to $\sim150\,h^{-1}{\rm Mpc}$. The residual panel~(f) displays the ratio $\xi_{\rm \textsc{phantom}}/\xi_{\rm \textsc{colossus}}$; for $r\lesssim80\,h^{-1}{\rm Mpc}$ the curves remain effectively indistinguishable from unity, while at larger radii the ratio oscillates between $\simeq0.985$ and $\simeq1.006$, corresponding to deviations of at most $\sim1.5\%$ as $\xi$ passes through zero. This behaviour is expected: once $|\xi_{\rm mm}|$ falls to $\sim10^{-4}$ and changes sign near $r\sim100\,h^{-1}{\rm Mpc}$, even tiny absolute differences between the codes translate into large fractional excursions, a limitation already noted for \textsc{colossus} \citep{Diemer_2018}. Panel~(g) shows the internal method comparison; here the direct integral and FFTLog implementations in \textsc{phantom} yield $\xi_{\rm int}(r)/\xi_{\rm fftlog}(r)=1$ at all sampled radii to machine precision, so the two approaches are numerically indistinguishable over the range plotted.

\section{Halo mass function and halo bias}
\label{sec:hmf_bias}

The halo mass function (HMF) connects the statistics of the linear density field to the abundance of collapsed structures. In \textsc{phantom}, we write the differential number density as
\begin{equation}
  \frac{\mathrm{d}n}{\mathrm{d}\ln M}
  = \frac{\rho_{\mathrm{m},0}}{M}
    \, f(\sigma)
    \left|\frac{\mathrm{d}\ln \sigma^{-1}}{\mathrm{d}\ln M}\right|,
  \label{eq:hmf_master}
\end{equation}
where $\rho_{\mathrm{m},0}$ is the present-day mean matter density, $\sigma(M,z)$ is the variance of the linear density field smoothed on the Lagrangian scale associated with mass $M$, and $f(\sigma)$ is the multiplicity function. All model dependence is absorbed into the choice of $f(\sigma)$ and into the collapse threshold $\delta_{\mathrm{c}}(z)$ provided by the cosmology module (Section~\ref{sec:cosmology}). The linear power spectrum, transfer-function suppression for WDM or FDM, and the choice of smoothing filter are handled by the machinery developed in Section~\ref{subsec:powerspec} and do not need to be specified manually in the HMF calls.

The simplest implementation in \textsc{phantom} is the Press--Schechter model, which assumes a constant spherical-collapse barrier and a sharp mass fraction above $\delta_{\mathrm{c}}$, leading to the familiar
$f(\sigma) \propto (\delta_{\mathrm{c}}/\sigma)\, \exp[-\delta_{\mathrm{c}}^{2}/(2\sigma^{2})]$ form. More flexible parametrisations such as Sheth--Tormen and its descendants replace this expression by a four-parameter fit in peak height $\nu \equiv \delta_{\mathrm{c}}/\sigma$ and calibrate the coefficients against $N$-body simulations.\footnote{The full list of implemented multiplicity functions, including their fiducial mass definitions and redshift ranges, is given in Table~\ref{tab:phantom_models_ph}.} \textsc{phantom} exposes these fits through a single dispatcher that evaluates the chosen $f(\sigma)$, differentiates the variance $\sigma^{2}(M,z)$ with respect to mass using the requested filter, and returns $\mathrm{d}n/\mathrm{d}\ln M$ on any user-supplied mass grid.

\begin{figure*}
  \centering
  \subfloat[\label{subfig:hmf_all_cdm}]{
    \includegraphics[width=0.32\textwidth]{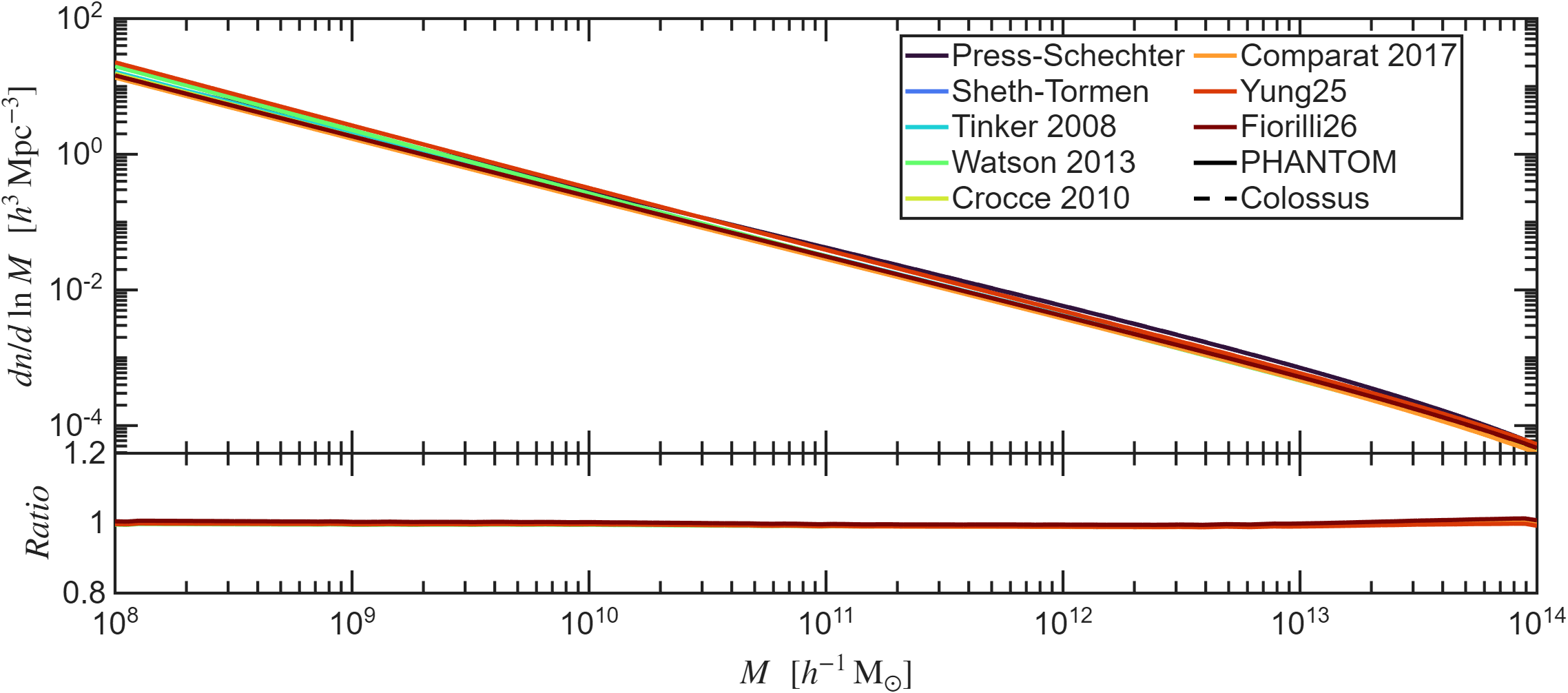}
  }
  \hfill
  \subfloat[\label{subfig:hmf_all_hmph}]{
    \includegraphics[width=0.32\textwidth, height=2.5cm]{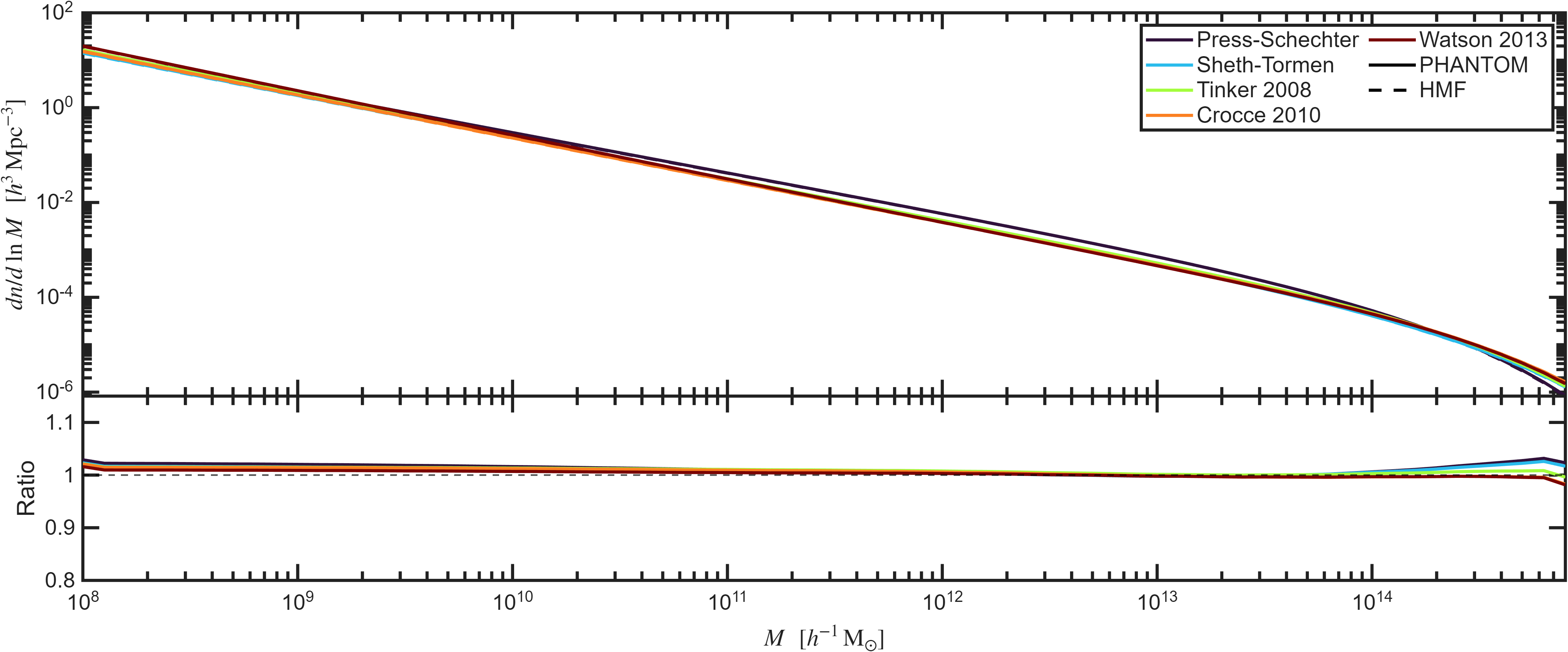}
  }
  \hfill
  \subfloat[\label{subfig:hmf_all_wdmfdm}]{
    \includegraphics[width=0.32\textwidth, height=2.5cm]{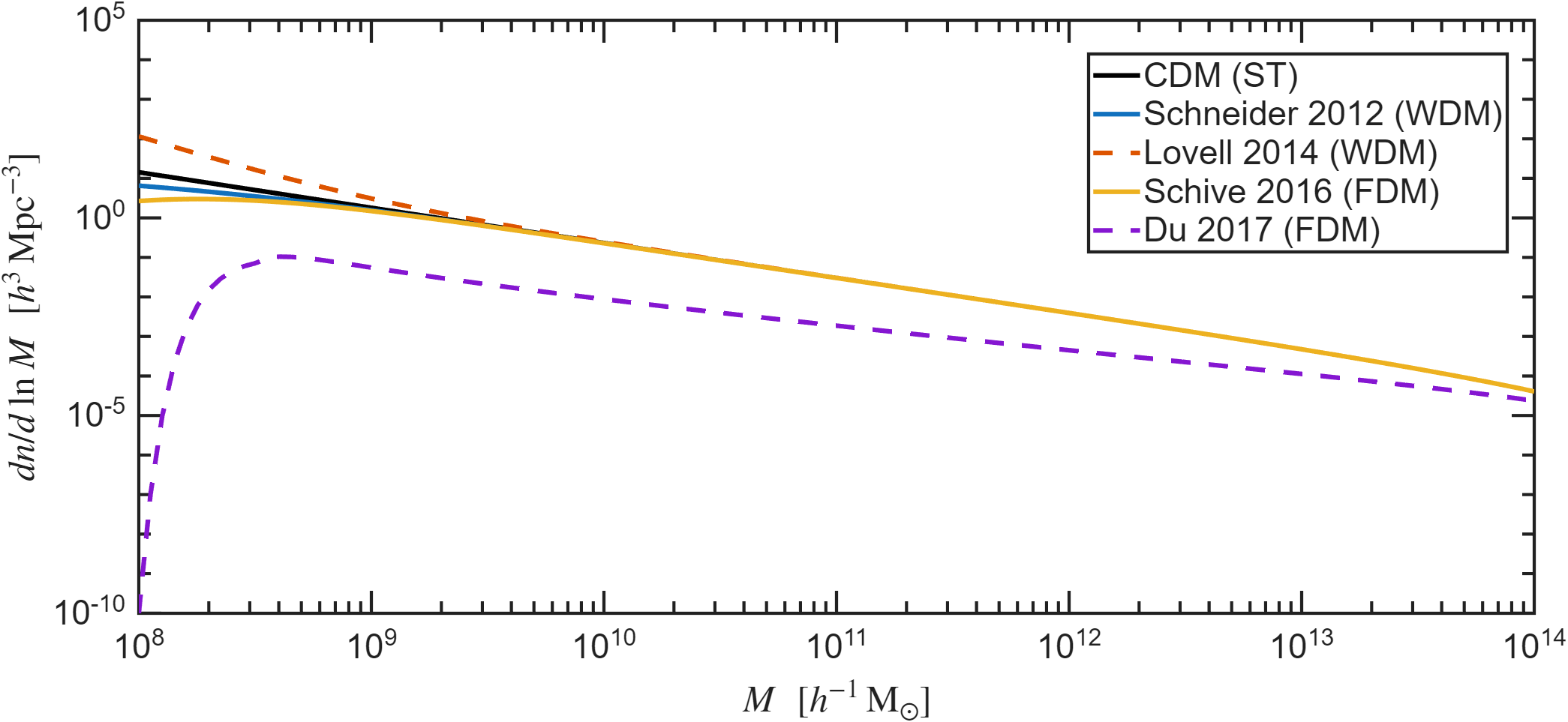}
  }
  \caption{Halo mass functions at $z=0$ computed with the models implemented in \textsc{phantom}. Panel~(\ref{subfig:hmf_all_cdm}) shows CDM predictions from several multiplicity functions; the lower sub-panel plots the ratio of each curve to \textsc{colossus} evaluated with the same cosmology and mass definition, demonstrating sub-percent agreement over $10^{8} \lesssim M / (h^{-1}\mathrm{M}_{\odot}) \lesssim 10^{14}$. Panel~(\ref{subfig:hmf_all_hmph}) compares the \textsc{phantom} CDM halo mass function against \textsc{hmf} \citep{Murray_2013}, a dedicated Python package for halo mass function calculations; since \textsc{halomod} \citep{Murray_2021} adopts \textsc{hmf} as its internal halo mass function engine, this comparison extends to \textsc{halomod} by construction. Panel~(\ref{subfig:hmf_all_wdmfdm}) displays the effect of non-cold dark matter models: warm dark matter (WDM) fits following~\citet{Schneider_2012} and~\citet{Lovel_2014}, and fuzzy dark matter (FDM) fits following~\citet{Schive_2016} and~\citet{Du_2016}. In all cases the same variance machinery is used, so differences are driven entirely by the underlying transfer functions and multiplicity calibrations.}
  \label{fig:hmf_all}
\end{figure*}

Figure~\ref{fig:hmf_all} collects the resulting halo mass functions at $z=0$ for a Planck-like cosmology. Panel~(\ref{subfig:hmf_all_cdm}) compares several CDM multiplicity models and shows the ratio with respect to \textsc{colossus} for identical input parameters. Across the halo mass range
$10^{8} \lesssim M / (h^{-1}\mathrm{M}_{\odot}) \lesssim 10^{14}$, the \textsc{phantom} implementations agree with \textsc{colossus} at the sub-percent level in both normalisation and slope, with residuals dominated by small differences in the numerical derivative of $\sigma(M)$ rather than by the multiplicity-function fits themselves. Panel~(\ref{subfig:hmf_all_hmph}) provides a direct comparison against \textsc{hmf} \citep{Murray_2013}, which uses the same formalism and multiplicity functions; because \textsc{halomod} \citep{Murray_2021} calls \textsc{hmf} internally for its halo mass function, agreement with \textsc{hmf} implies consistency with \textsc{halomod} as well.

For non-cold dark matter models, the linear power spectrum is modified before constructing $\sigma(M,z)$, and the same machinery is used to obtain the suppressed HMF. Panel~(\ref{subfig:hmf_all_wdmfdm}) illustrates this for a representative WDM thermal relic and an FDM model with boson mass $m_{22} \equiv m/(10^{-22}\,\mathrm{eV})$ held fixed. The WDM HMFs of \citet{Schneider_2012} and \citet{Lovel_2014} follow the CDM prediction at high masses and develop a smooth cutoff below the half-mode mass $M_{1/2}$, while the FDM implementations of \citet{Schive_2016} and \citet{Du_2016} produce a more abrupt suppression linked to the Jeans scale. All curves are obtained by feeding the corresponding suppressed transfer functions into the variance calculation, so the WDM and FDM HMFs remain fully consistent with the power-spectrum and filter choices used elsewhere in the code.

The halo bias module provides the linear response of halo number density to long-wavelength background perturbations and is evaluated as a function of $\sigma(M,z)$ and $\delta_{\mathrm{c}}$ as well. For a given bias model $b(\nu)$, the dispatcher takes as input either $\sigma$ or $M$ and returns the scale-independent Eulerian bias factor $b(M,z)$; angle-averaged two-halo terms and large-scale clustering predictions are then built from this quantity in subsequent analysis. The simplest option, corresponding to the Cole--Kaiser relation, reads $b(\nu) = 1 + (\nu^{2}-1)/\delta_{\mathrm{c}}$ and is included mainly for comparison. More accurate calibrations, such as \citet{Jing_1998},\citet{Sheth_1999}, \citet{Sheth_2001}, \citet{Seljak_2004}, \citet{Tinker_2010}, \citet{Pillepich_2010}, \citet{Bhattacharya_2011}, and \citet{Comparat_2017}, are exposed through dedicated bias functions that follow the original fitting forms and parameter ranges, with \citet{Tinker_2010} used as the default CDM choice.

Beyond-CDM models are handled in an implicit fashion: for WDM, the \citet{Schneider_2012} prescription applies the \citet{Sheth_2001} bias formula to the WDM variance $\sigma(M,z)$ and masks the spurious halo regime below the half-mode mass using the half-mode mass helper functions. For FDM, the collapse barrier is mass-dependent because quantum pressure suppresses growth below the Jeans mass $M_{\rm J}$. Following \citet{Marsh_2014} and \citet{Du_2016}, the effective collapse threshold is
\begin{equation}
    \delta_c^{\rm fdm}(M, z) = G(M)\,\delta_c^{\rm cdm}(z),
    \label{eq:delta_c_fdm}
\end{equation}
where $G(M) = D_{\rm cdm}(z)/D_{\rm fdm}(M,z)$ is the ratio of the CDM growth factor to the suppressed FDM growth factor at mass scale $M$, and $G(M) \rightarrow 1$ for $M \gg M_{\rm J}$ (recovering the CDM limit). The fitting function for $G(M)$ from \citet{Marsh_2016} is implemented in collapse overdensity for FDM and used by the multiplicity function of \texttt{du17}. The bias for FDM is computed via the Sheth--Mo--Tormen moving-barrier model \citep[\texttt{smt01};][]{Sheth_2001}, with $\delta_c^{\rm fdm}(M,z)$ from Eq.~\eqref{eq:delta_c_fdm} substituted in place of the CDM value. This model is the natural companion to the \texttt{du17} multiplicity function: both originate from the same moving-barrier excursion-set framework \citep{Marsh_2014, Du_2016}, so the bias reshapes self-consistently when the barrier is modified, without requiring a separately calibrated fit. A runtime warning is issued whenever the bias is evaluated within a decade in mass of $M_{\rm J}$, since the barrier calibration is derived for $M \gg M_{\rm J}$ and this regime remains unconstrained by dedicated FDM N-body runs. A complete list of available bias models is given in Table~\ref{tab:phantom_models_ph}.

\section{Halo Observables}
\label{sec:halo_obs}

Dark matter haloes are characterised by a set of structural and kinematic quantities that connect the density field to observable predictions. \textsc{phantom} provides routines for computing these quantities self-consistently given a halo virial mass $M_h$, redshift $z$, and cosmology. The following two subsections describe the general halo properties that hold regardless of the dark matter model, and the additional observables that are specific
to Fuzzy Dark Matter (FDM) haloes.

\subsection{General Halo Properties}
\label{subsec:general_halo}

Following the spherical overdensity convention \citep{Bryan_1998}, the virial radius $R_\Delta$ is defined as the radius enclosing a mean interior density equal to $\Delta\,\rho_\mathrm{ref}(z)$, where $\rho_\mathrm{ref}$ is either the critical or mean matter density of the universe at redshift $z$:
\begin{equation}
    M_\Delta = \frac{4}{3}\pi\,R_\Delta^3\,
    \Delta\,\rho_\mathrm{ref}(z).
\label{eq:SO_def}
\end{equation}
\textsc{phantom} evaluates $\rho_\mathrm{ref}(z)$ through the cosmology module, which computes the Hubble parameter $E(z)$ exactly and returns $\rho_\mathrm{crit}(z) = \rho_{\mathrm{crit},0}\,E^2(z)$. The virial overdensity $\Delta_\mathrm{vir}(z)$ is obtained from the fitting formula of \citet{Bryan_1998}. From the enclosed mass ($M_{\Delta}$) and the virial radius ($R_{\Delta}$), the virial circular velocity ($v_{\Delta}$) follows directly:
\begin{equation}
    V_\Delta = \sqrt{\frac{G\,M_\Delta}{R_\Delta}}.
\label{eq:V_vir}
\end{equation}
The dynamical time ($t_{dyn}$), the time for a test particle to cross the
halo at the virial velocity, is~\citep{Diemer_2018}
\begin{equation}
    t_\mathrm{dyn}(z) = \frac{2\,R_\Delta}{V_\Delta}
    = \sqrt{\frac{3}{4\pi G\,\Delta\,\rho_\mathrm{ref}(z)}},
\label{eq:t_dyn}
\end{equation}
which depends only on the mean interior density and not on the distribution of mass within the halo. Eq.~\eqref{eq:t_dyn} is equivalent to the Hubble time $t_H(z) = [H(z)]^{-1}$ up to factors of order unity, reflecting the fact that virialized structures collapse on a timescale set by the ambient expansion rate.


The outer halo follows the Navarro--Frenk--White (NFW) density profile \citep{Navarro_1997}, whose functional form is discussed in Section~\ref{subsec:density_profiles}. The profile is parameterised by a scale density $\rho_s$ and a scale radius $r_s$; how these are determined from the concentration--mass relation is described in Section~\ref{subsec:concentration}. From the NFW profile, \textsc{phantom} computes the enclosed mass $M_\mathrm{NFW}(r)$, the circular velocity $V_c(r)$, and the maximum circular velocity $V_\mathrm{max}$, which occurs at $r_\mathrm{max} \approx 2.16\,r_s$~\citep{Navarro_1997}. The projected surface density and the line-of-sight velocity dispersion are obtained from the three-dimensional profiles via Abel integrals. The surface mass density at projected radius $R$ is
\begin{equation}
    \Sigma(R) = 2\int_R^{\infty}
    \frac{\rho(r)\,r}{\sqrt{r^2 - R^2}}\,\mathrm{d}r,
\label{eq:Sigma}
\end{equation}
and the mass-weighted line-of-sight velocity dispersion $\sigma_{\rm los}(R)$, obtained by projecting the isotropic Jeans equation along the line of sight for a spherically symmetric dark matter halo of density $\rho(r)$ \citep{Binney_2008}:
\begin{equation}
    \sigma_\mathrm{los}^2(R)
    = \frac{2}{\Sigma(R)}\int_R^{\infty}
    \left(1 - \frac{R^2}{r^2}\right)
    \rho\,\sigma_r^2\,\frac{r}{\sqrt{r^2 - R^2}}\,\mathrm{d}r,
\label{eq:sigma_los}
\end{equation}
where $\sigma_r(r)$ is the three-dimensional radial velocity dispersion obtained from the isotropic Jeans equation~\citep{Binney_2008}:
\begin{equation}
    \sigma_r^2(r) = \frac{1}{\rho(r)}
    \int_r^{\infty} \rho(r')\,
    \frac{G\,M(r')}{r'^2}\,\mathrm{d}r'.
\label{eq:sigma_r}
\end{equation}
The integrals in Eqs.~\eqref{eq:Sigma}--\eqref{eq:sigma_r} are evaluated numerically on a logarithmically spaced radial grid using a singularity-free substitution that removes the integrable divergence at $r = R$ in Eqs.~\eqref{eq:Sigma} and \eqref{eq:sigma_los}.


\subsection{FDM Halo Observables}
\label{subsec:fdm_obs}

FDM haloes are distinguished from their cold dark matter counterparts by the presence of a gravitationally self-bound solitonic core at their centre. The core forms as the ground-state solution of the Schr\"{o}dinger--Poisson equations and is embedded within an outer halo whose profile asymptotes to the NFW form at large radii. The soliton density profile and its numerical derivation are discussed in Section~\ref{subsec:density_profiles}. Unless otherwise noted, all scaling relations in this subsection follow \citet{Schieve_2014_core}, \citet{Schive_2014_cosmic}, \citet{Robles_2018}, \citet{Mocz_2017}, and \citet{Chan_2022}.

The peak central density $\rho_c$ and the core radius $r_c$ — defined as the radius at which the density drops to one-half its peak value i.e. $\rho(r_c) = \frac{\rho_c}{2}$ — are related by the scaling symmetry of the Schr\"{o}dinger--Poisson equations:
\begin{equation}
    \rho_c = 1.93 \times 10^{7}\,
    m_{22}^{2}\,
    \left(\frac{r_c}{\mathrm{kpc}}\right)^{-4}
    \; M_\odot\,\mathrm{kpc}^{-3},
\label{eq:rhoc_rc}
\end{equation}
where $m_{22} \equiv m / (10^{-22}\,\mathrm{eV})$ is the boson mass in units of $10^{-22}\,\mathrm{eV}$. Eq.~\eqref{eq:rhoc_rc} is invertible: specifying either $\rho_c$ or $r_c$ fully determines the other. The core mass, defined as $M_c = M(r \leq r_c)$, follows from the soliton scaling relation:
\begin{equation}
    M_c = \frac{5.04 \times 10^{7}}{m_{22}}
    \left(\frac{M_h}{10^9 M_{\odot}}\right)^{1/3}
    \; M_\odot.
\label{eq:Mc_rc}
\end{equation}
The product $r_c M_c$ is independent of halo mass:
\begin{equation}
    r_c\,M_c = \frac{8.06\times 10^{7}}{m_{22}^{2}}
    \; M_\odot\,\mathrm{kpc},
\label{eq:rcMc}
\end{equation}
which, through $j_c = \sqrt{G\,r_c\,M_c}$, implies that the specific angular momentum of a circular orbit within the soliton depends only on the boson mass:
\begin{equation}
    j_c \approx \frac{18.6}{m_{22}}
    \; \mathrm{kpc}\,\mathrm{km}\,\mathrm{s}^{-1}.
\label{eq:jc}
\end{equation}

The core-halo mass relation (CHMR) connects the soliton size to the virial mass of the host halo. Cosmological simulations show that the core radius scales with halo mass and scale factor $a$ as
\begin{equation}
    r_c = 1.6\,m_{22}^{-1}\,a^{1/2}
    \left(\frac{M_h}{10^{9}\,M_\odot}\right)^{-1/3}
    \; \mathrm{kpc},
\label{eq:rc_Mh}
\end{equation}
 where $a$ is the cosmic scale factor. The scaling $r_c \propto a^{1/2}$ follows \citet{Schieve_2014_core}, where it was calibrated from cosmological simulations spanning a limited halo mass range at $z \lesssim 1$; its accuracy at higher redshifts and for cluster-mass haloes remains to be confirmed with larger-volume FDM simulations. This relation is invertible so that either the halo mass or the core radius can serve as the input quantity. Together with Eq.~\eqref{eq:Mc_rc}, the CHMR implies that the soliton core mass scales as $M_c \propto M_h^{1/3}$, whereas the core mass fraction decreases with halo mass:
\begin{equation}
    \frac{M_c}{M_h}
    = 5.04\times10^{-2}
    \left(\frac{M_h}{10^{9}\,M_\odot}\right)^{-2/3}
    m_{22}^{-1},
\label{eq:McMh}
\end{equation}
and the ratio of the core to virial radius likewise decreases:
\begin{equation}
    \frac{r_c}{R_\mathrm{vir}}
    = 6.20\times10^{-2}
    \left(\frac{M_h}{10^{9}\,M_\odot}\right)^{-2/3}
    m_{22}^{-1}.
\label{eq:rcRvir}
\end{equation}
Eqs.~\eqref{eq:McMh} and \eqref{eq:rcRvir} show that the soliton dominates a progressively smaller fraction of the halo in more massive systems. The circular velocity within the soliton reaches its maximum near $r_c$. Together with the NFW profile of the outer halo, \textsc{phantom} constructs the composite circular velocity curve $V_c(r) = \sqrt{G M(r)/r}$ across the full radial range. Following \citet{Robles_2018}, the ratio $\frac{V_{\rm c}(r_c)}{V_{\rm c}(r_{\rm vir}}) \approx 0.9$ holds independently of both halo mass and boson mass, and serves here as a self-consistency check on the composite profile normalisation. The surface density $\Sigma(R)$ and line-of-sight velocity dispersion $\sigma_\mathrm{los}(R)$ are computed from the full composite density profile using Eqs.~\eqref{eq:Sigma}--\eqref{eq:sigma_r}, enabling direct comparison with kinematic and lensing observations of FDM halo candidates.

\subsection{Density Profiles}
\label{subsec:density_profiles}

Dark matter halo density profiles encode the radial mass distribution from the innermost resolved region out to the virial boundary. \textsc{phantom} supports a suite of spherically symmetric profiles that span from classical CDM parametrisations to the quantum core structure of FDM haloes; the full list of implemented models and their dispatcher keys is given in Table~\ref{tab:phantom_models_dc}. In all cases, the profile is normalised by enforcing mass conservation within $R_\Delta$, and the enclosed mass $M(r)$, circular velocity $V_\mathrm{c}(r)$, and surface density $\Sigma(R)$ follow from the profile analytically or by numerical quadrature.

\paragraph{Navarro--Frenk--White profile.}
The NFW profile \citep{Navarro_1997} is the canonical result of collisionless CDM $N$-body simulations and remains the reference against which all alternative models are compared. It is characterised by a central density cusp scaling as $\rho \propto r^{-1}$ and an outer fall-off as $r^{-3}$:
\begin{equation}
    \rho_\mathrm{NFW}(r) =
    \frac{\rho_s}{\left(r/r_s\right)
    \left(1 + r/r_s\right)^{2}},
\label{eq:NFW}
\end{equation}
where $\rho_s$ is the characteristic scale density and $r_s$ is the scale radius. The enclosed mass follows analytically as %
\begin{equation}
    M_\mathrm{NFW}(r) = 4\pi\,\rho_s\,r_s^3
    \left[\ln\!\left(1 + \frac{r}{r_s}\right)
    - \frac{r/r_s}{1 + r/r_s}\right].
\label{eq:M_NFW}
\end{equation}
Given $M_\Delta$, $R_\Delta$, and concentration $c \equiv R_\Delta/r_s$ (Section~\ref{subsec:concentration}), the scale  parameters are fixed by $r_s = R_\Delta/c$ and mass conservation, yielding~\citep{Navarro_1997}
\begin{equation}
    \rho_s = \frac{M_\Delta}{4\pi\,r_s^3\,
    \left[\ln(1+c) - c/(1+c)\right]}.
\label{eq:rhos}
\end{equation}
The circular velocity reaches its maximum at $r_\mathrm{max} \approx 2.16\,r_s$ \citep{Navarro_1997}.

\paragraph{Hernquist profile.}
The Hernquist profile \citep{Hernquist_1990} approximates the de Vaucouleurs $r^{1/4}$ surface brightness law of elliptical galaxies while retaining a closed-form three-dimensional density:
\begin{equation}
    \rho_\mathrm{H}(r) =
    \frac{\rho_s}{\left(r/r_s\right)
    \left(1 + r/r_s\right)^{3}}.
\label{eq:Hernquist}
\end{equation}
The outer slope steepens from $r^{-3}$ to $r^{-4}$ relative to the NFW form, so the total mass converges to the finite value $M_\mathrm{tot} = 2\pi\,\rho_s\,r_s^3$. The normalisation is set by $M_\mathrm{tot} = M_\Delta\,(1+c)^2/c^2$ \citep{Hernquist_1990}.

\paragraph{Einasto profile.}
The Einasto profile \citep{Einasto_1965} replaces the power-law inner cusp with a smooth exponential, providing a closer match to the inner structure seen in high-resolution CDM simulations \citep{Merritt_2006}:
\begin{equation}
    \rho_\mathrm{Ein}(r) = \rho_s\,
    \exp\!\left\{-\frac{2}{\alpha_e}
    \left[\left(\frac{r}{r_s}\right)^{\alpha_e} - 1\right]
    \right\}.
\label{eq:Einasto}
\end{equation}
The shape parameter $\alpha_e$ controls the profile curvature; smaller values produce a more concentrated, cusp-like interior. Rather than treating $\alpha_e$ as a free parameter, it is determined from the dimensionless peak height $\nu(M,z) \equiv \delta_c(z)/\sigma(M,z)$ via the empirical calibration of \citet{Gao_2008}:
\begin{equation}
    \alpha_e = 0.155 + 0.0095\,\nu^2,
\label{eq:alpha_e}
\end{equation}
capped at $\alpha_e = 0.3$ following \citet{Benson_2011}, beyond which the \citet{Gao_2008} calibration was not constrained. Low-mass, common haloes ($\nu \lesssim 1$) have $\alpha_e \approx 0.16$; rare, massive clusters ($\nu \sim 3$) reach $\alpha_e \approx 0.24$--$0.30$.

\paragraph{Diemer--Kravtsov profile.}
The \citet{Diemer_2014} profile extends the Einasto form to explicitly model the splashback feature — the caustic-like density drop near the outermost apocenter of recently accreted material — and adds a power-law term for the infalling outer envelope:
\begin{equation}
    \rho_\mathrm{DK14}(r) =
    \rho_\mathrm{Ein}(r)\,f_\mathrm{trans}(r)
    + \rho_\mathrm{out}(r).
\label{eq:DK14}
\end{equation}
The truncation function
\begin{equation}
    f_\mathrm{trans}(r) =
    \left[1 + \left(\frac{r}{r_t}\right)^\beta\right]^{-\gamma_t/\beta}
\label{eq:f_trans}
\end{equation}
equals unity for $r \ll r_t$ and suppresses the density sharply beyond the splashback radius $r_t$. For a mass-selected sample, $(\beta,\gamma_t) = (4,8)$ and \citep{Diemer_2014}
\begin{equation}
    r_t = R_{200m}\left(1.9 - 0.18\,\nu_{200m}\right),
\label{eq:rt}
\end{equation}
For samples selected jointly by mass and mass accretion rate $\Gamma$, the parameters shift to $(\beta,\gamma_t) = (6,4)$ with $r_t = 0.54\,R_{200m}(1 +
0.53\,e^{-\Gamma})$. The outer infalling term is
\begin{equation}
    \rho_\mathrm{out}(r) = \bar{\rho}_m(z)\,
    \left(\frac{r}{r_\mathrm{ref}}\right)^{-s_e},
\label{eq:rho_out}
\end{equation}
with $s_e = 1.5$ and $r_\mathrm{ref} = 1\,\mathrm{Mpc}\,h^{-1}$ \citep{Diemer_2014}. This is the only profile in the suite that is physically motivated beyond the virial boundary and captures the mean density of the surrounding environment. In \textsc{phantom}, $r_t$ is evaluated using $\nu_{200m}$ computed at $M_{200m}$, which is obtained from $M_{200c}$ via an NFW spherical overdensity conversion at fixed concentration and redshift.

\begin{figure}
    \centering
    \includegraphics[width=\columnwidth]{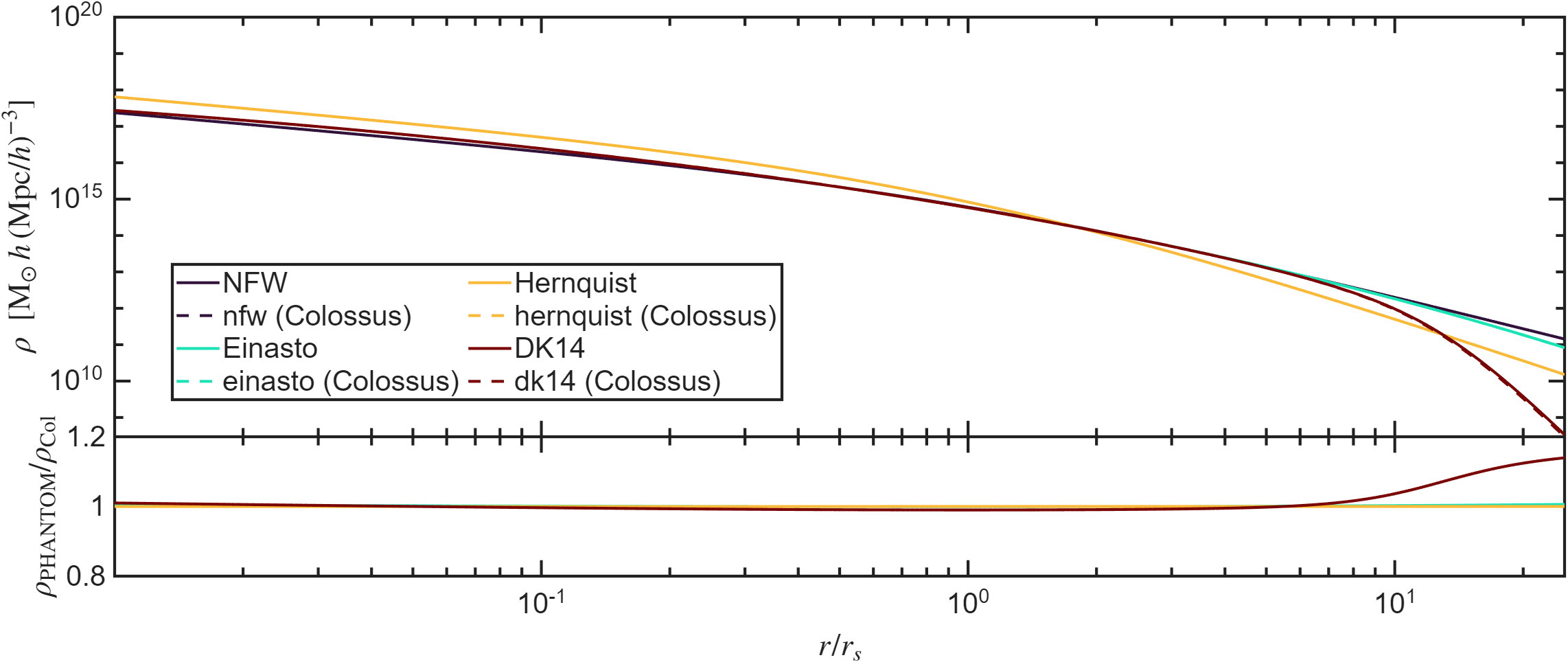}
    \caption{CDM density profiles computed by \textsc{phantom} (solid lines) compared to \textsc{colossus} \citep{Diemer_2018} (dashed lines) for a halo of    $M_{200c} = 10^{14}\,\mathrm{M_\odot}\,h^{-1}$, $c = 5$, $z = 0$. The lower panel shows the ratio $\rho_\mathrm{PHANTOM}/\rho_\mathrm{Col}$. NFW and Hernquist agree at the $0.1$ per cent level across the full radial range. The Einasto profile shows a constant $\sim\!0.5$ per cent offset arising from the peak-height mass-definition convention (Section~\ref{subsec:density_profiles}). The DK14 deviation beyond $r \gtrsim 10\,r_s$ is driven by the exponential sensitivity of the splashback truncation term and lies within the calibration scatter of \citet{Diemer_2014}.}
    \label{fig:profile_CDM}
\end{figure}

Figure~\ref{fig:profile_CDM} shows all four CDM profiles computed by \textsc{phantom} for a cluster-mass halo of $M_{200c} = 10^{14}\,\mathrm{M_\odot}\,h^{-1}$, concentration $c = 5$, and $z = 0$, compared against the independent Python implementation in \textsc{colossus} \citep{Diemer_2018}. The NFW and Hernquist profiles agree with \textsc{colossus} to better than $0.1$ per cent across the full radial range. The Einasto profile matches at the $0.5$ per cent level; the small residual traces the difference between the virial peak height $\nu_\mathrm{vir}$ used by \textsc{colossus} to evaluate Eq.~\eqref{eq:alpha_e} and the $200c$ peak height used in earlier versions of the code. In \textsc{phantom}, $\alpha_e$ is computed at the virial mass $M_\mathrm{vir}$ obtained by converting $M_{200c}$ to the virial mass definition via an NFW-based spherical overdensity conversion. The DK14 profile agrees with \textsc{colossus} to within $1$ per cent for $r \lesssim R_{200c}$. The deviation beyond that radius reflects the sensitivity of the splashback truncation term (Eq.~\eqref{eq:f_trans}) to the peak height used in Eq.~(\eqref{eq:rt}): the truncation function falls steeply as $f_\mathrm{trans} \propto (r/r_t)^{-\gamma_t/\beta}$, so a sub-per-cent difference in $\nu_{200m}$ between the two codes is amplified exponentially at $r \gg r_t$. This behaviour is within the intrinsic scatter of the \citet{Diemer_2014} calibration, which reports 10--20 per cent dispersion in $r_t$ at fixed $\nu_{200m}$.

\paragraph{Soliton and composite FDM profile.}
FDM haloes require a fundamentally different description at small radii. The ground-state solution of the Schr\"{o}dinger--Poisson equations produces a gravitationally self-bound solitonic core whose density profile is well approximated by \citep{Schive_2014_cosmic, Mocz_2017, Hui_2017, Robles_2018}
\begin{equation}
    \rho_\mathrm{sol}(r) =
    \frac{\rho_c}{\left[1 + 0.091\,
    (r/r_c)^2\right]^{8}},
\label{eq:soliton_profile}
\end{equation}
where $\rho_c$ is the peak central density and $r_c$ is the core radius at which the density drops to half of its peak value. The profile has a near-constant-density interior for $r \ll r_c$ and a steep outer gradient, and is entirely distinct from cores produced by warm or self-interacting dark matter, both of which merely truncate an NFW cusp rather than replace it with a self-bound quantum object \citep{Schieve_2014_core}. Cosmological FDM simulations show that the solitonic core is surrounded by a granular halo whose azimuthally averaged density asymptotes to the NFW form at large radii
\citep{Schive_2014_cosmic,Schieve_2014_core,Mocz_2017}. The full halo
profile is therefore modelled as a composite:
\begin{equation}
    \rho(r) =
    \begin{cases}
        \rho_\mathrm{sol}(r) & r \leq r_x, \\
        \rho_\mathrm{NFW}(r) & r > r_x,
    \end{cases}
\label{eq:composite}
\end{equation}
where $r_x$ is the transition radius defined by $\rho_\mathrm{sol}(r_x) = \rho_\mathrm{NFW}(r_x)$~\citep{Chan_2022, Robles_2018}. Simulations find $r_x \approx \xi\,r_c$ with $\xi \approx 2$--$4$, increasing mildly with halo mass \citep{Robles_2018}. This composite can be constructed analytically, with $\rho_c$ and $r_c$ set by the core-halo mass relation (Section~\ref{subsec:fdm_obs}) and the NFW parameters determined from $M_\mathrm{vir}$, $R_\mathrm{vir}$, and $c$; or it can be built by fitting both components directly to a spherically averaged simulation density profile, where the fitting region for each component and the selection of the best cutoff radius are optimised by a goodness-of-fit statistic. Both routes produce the same composite form and the quantities $\rho_c$, $r_c$, $\rho_s$, and $r_s$ are carried forward to all downstream observables described in Section~\ref{subsec:fdm_obs}.

\begin{figure}
    \centering
    \includegraphics[width=\columnwidth]{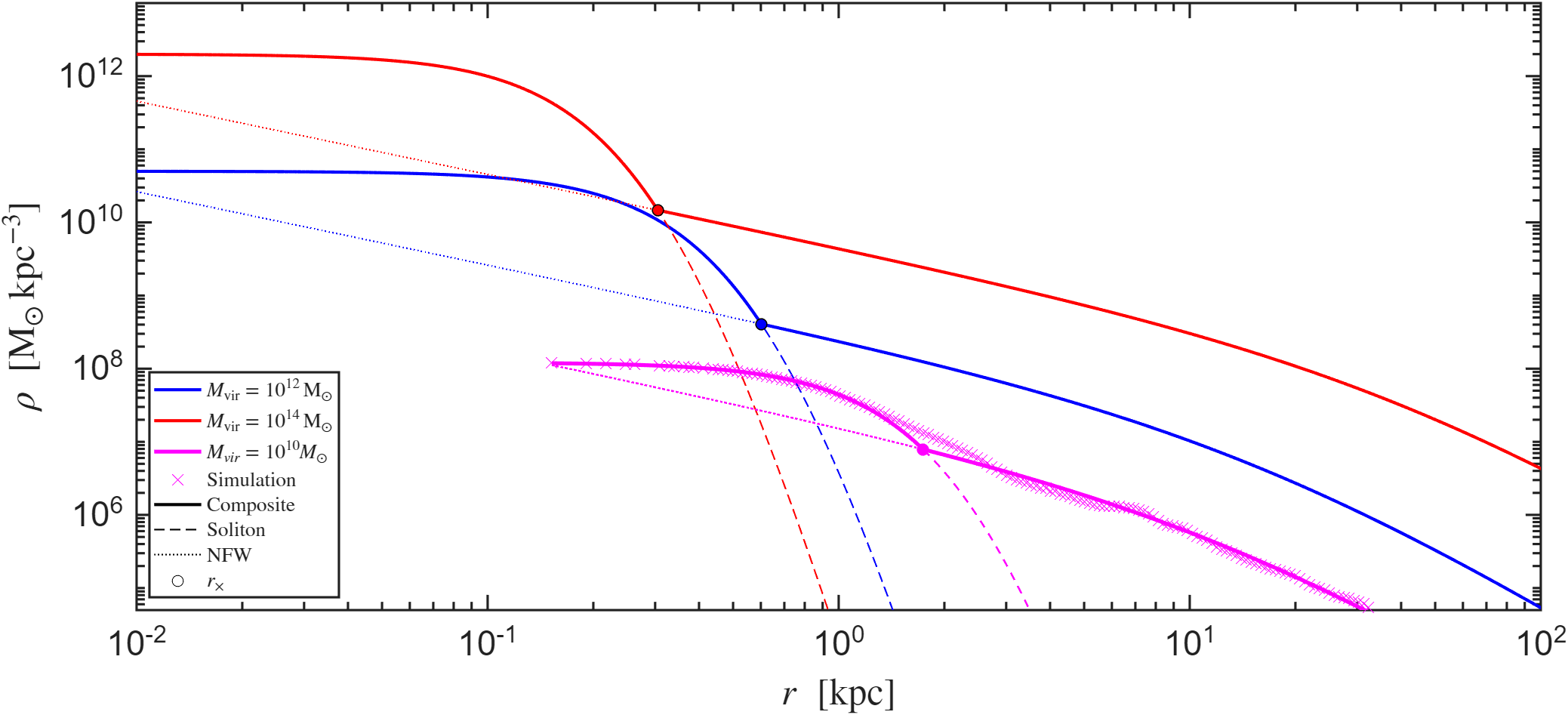}
    \caption{Composite FDM density profiles computed by \textsc{phantom} at $z = 0$ for three halo masses: $M_\mathrm{vir} \in \{10^{10},\,10^{12},\,10^{14}\}\,\mathrm{M_\odot}$ (magenta, blue, and red, respectively). For the analytic profiles (blue, red), solid lines show the full composite $\rho(r)$; dashed lines show the soliton component $\rho_\mathrm{sol}$; dotted lines show the NFW outer halo $\rho_\mathrm{NFW}$. For the dwarf-scale case ($M_\mathrm{vir} = 10^{10}\,\mathrm{M_\odot}$, magenta), cross markers show the radially binned density profile extracted from a dark-matter-only simulation conducted following \citet{Mocz_2017}, and the solid magenta line shows the composite profile fitted to these data by \textsc{phantom}; the dashed and dotted magenta lines show the corresponding best-fit soliton and NFW components, respectively (see Appendix~\ref{app:fitting} for the fitting procedure). Filled circles mark the transition radius $r_\times$ where $\rho_\mathrm{sol}(r_\times) = \rho_\mathrm{NFW}(r_\times)$ for each case. Analytic profiles adopt $m_\psi = 10^{-22}\,\mathrm{eV}$; the simulation uses $m_\psi = 0.5\times10^{-22}\,\mathrm{eV}$.}
    \label{fig:profile_FDM}
\end{figure}

Figure~\ref{fig:profile_FDM} shows the composite FDM density profile for three halo masses spanning four decades in $M_\mathrm{vir}$, computed analytically with \textsc{phantom} (blue and red) and fitted to a dark-matter-only simulation (magenta). The simulation follows the spectral Schr\"{o}dinger--Poisson solver of \citet{Mocz_2017} and produces a single virialized halo at $M_\mathrm{vir} = 10^{10}\,\mathrm{M_\odot}$ with $m_\psi = 0.5\times10^{-22}\,\mathrm{eV}$; \textsc{phantom} then fits the composite profile (Eq.~\ref{eq:composite}) to the radially binned density data, recovering the soliton and NFW components shown in the figure (see Appendix~\ref{app:fitting} for the fitting procedure and goodness-of-fit assessment). For each case, the transition radius $r_\times$ shifts outward with increasing halo mass, consistent with the core-halo mass relation discussed in Section~\ref{subsec:fdm_obs}: the dwarf-scale halo ($10^{10}\,\mathrm{M_\odot}$) is soliton-dominated out to $\sim\!0.5\,\mathrm{kpc}$, whereas the cluster-scale halo ($10^{14}\,\mathrm{M_\odot}$) transitions near $\sim\!0.3\,\mathrm{kpc}$ at a central density roughly eight orders of magnitude higher, reflecting the much deeper potential well of the host. The composite is continuous at $r_\times$ by construction; no smoothing kernel is applied, since both components return the same density at that radius by definition.

\subsection{Concentration--Mass Relation}
\label{subsec:concentration}

\begin{figure*}
  \centering
  \includegraphics[width=0.48\textwidth]{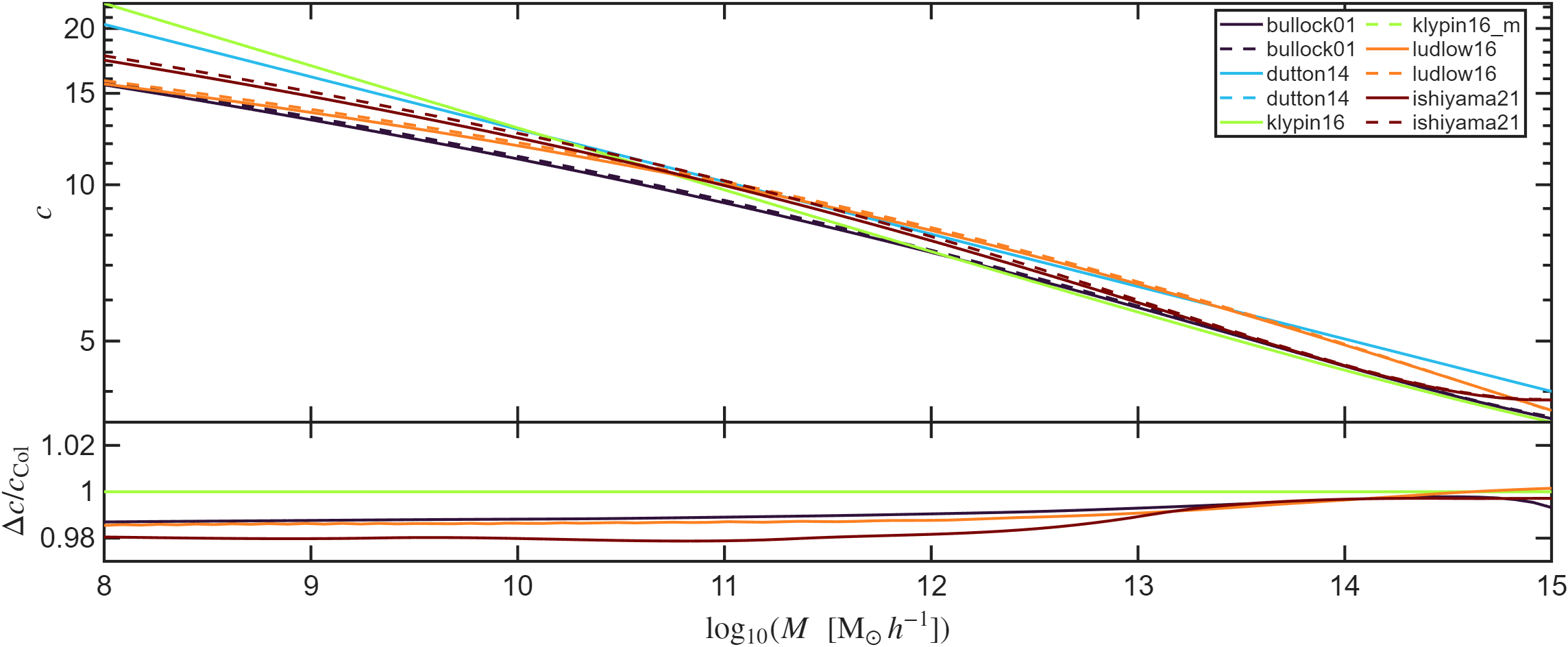}
  \label{subfig:c_comp}
  \hfill
  \includegraphics[width=0.48\textwidth]{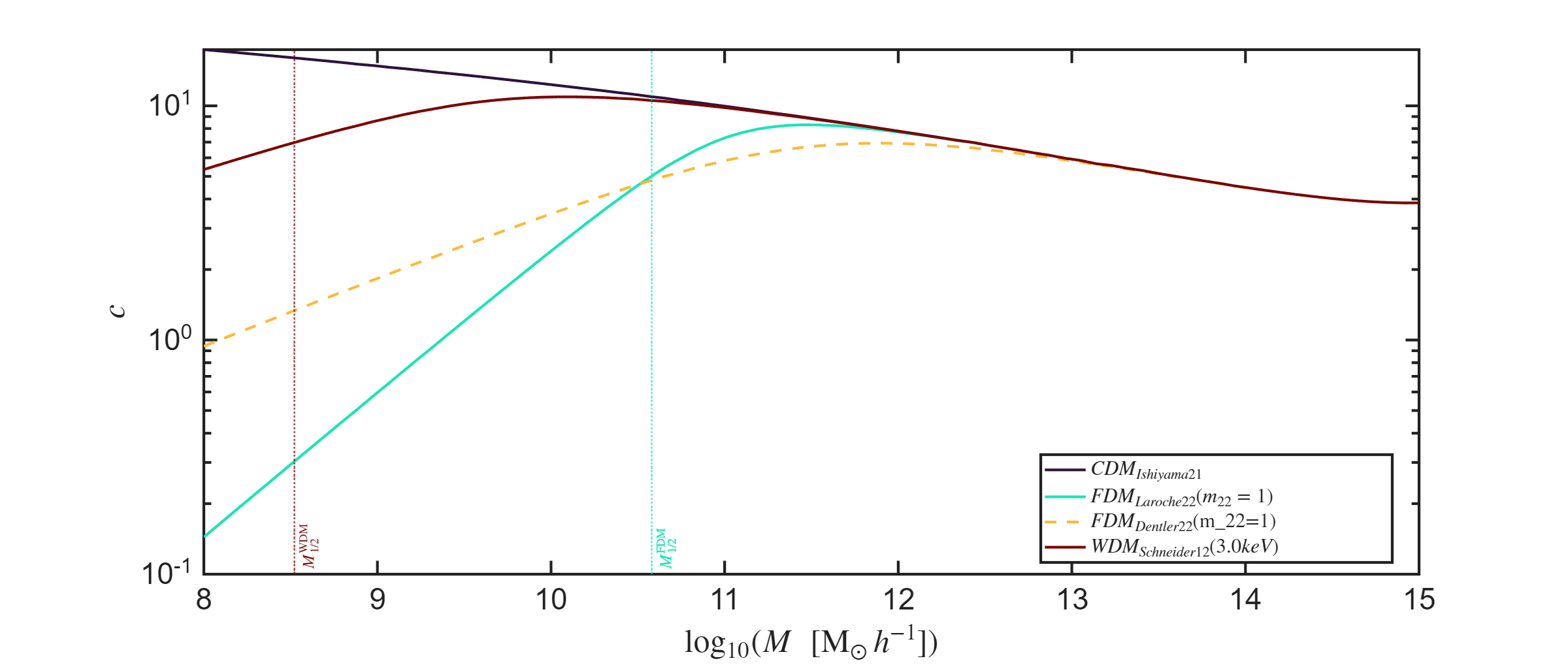}
  \label{subfig:c_sup}
  \caption{Concentration--mass relations at $z=0$ from \textsc{phantom}. \textit{Panel (a):} Five representative CDM models compared against \textsc{colossus} \citep{Diemer_2018}. Solid lines show \textsc{phantom} outputs; dashed lines of matching colour show the corresponding \textsc{colossus} predictions for \texttt{bullock01} \citep{Bullock_2001}, \texttt{dutton14} \citep{Dutton_2014}, \texttt{klypin16} \citep{klypin_2016}, \texttt{ludlow16} \citep{Ludlow_2016}, and \texttt{ishiyama21} \citep{Ishiyama_2021}. The upper sub-panel shows $c$ directly; the lower sub-panel shows the fractional residual $\Delta c/c_{\rm Col}$ relative to \textsc{colossus}, with agreement better than $1$--$2\%$ across $10^{8}$--$10^{15}\,h^{-1}M_\odot$. The remaining CDM models are implemented but omitted from the figure for clarity; all pass the same validation threshold. \textit{Panel (b):} Beyond-CDM models. The CDM baseline (\texttt{ishiyama21}, dark blue) is shown for reference. WDM at $3.0\,{\rm keV}$ (\texttt{schneider12}, dark red) and FDM at $m_{22}=1$ (\texttt{laroche22}, cyan solid; \texttt{dentler22}, orange dashed) all converge to CDM above their respective half-mode masses (vertical dotted lines, $M_{1/2}^{\rm WDM}$ in dark red, $M_{1/2}^{\rm FDM}$ in cyan). Below $M_{1/2}$, the \citet{Schneider_2012} WDM model drops steeply but remains finite, while the \citet{Laroche_2022} FDM model falls below $c=1$, indicating unconcentrated haloes where $r_s \geq r_{\rm vir}$. The \citet{Dentler_2022} prescription predicts a shallower decline in the same regime, reflecting differences in simulation calibration between the two FDM approaches.}
  \label{fig:conc}
\end{figure*}

The concentration--mass relation connects a halo's structural compactness to its assembly history: haloes that collapse earlier, when the mean cosmic density is higher, are more concentrated \citep{Wechsler_2002, Ishiyama_2021}. \textsc{phantom} implements twelve CDM models and three beyond-CDM models, all accessed through a unified dispatcher by specifying a string key; dark-matter--specific parameters (e.g.\ \texttt{cosmo.m22} or \texttt{cosmo.m\_wdm\_keV}) are set in the cosmology structure. The full list is given in Table~\ref{tab:phantom_models_dc}; detailed parameter descriptions and usage examples are provided in the \phantomwiki.

The CDM models span three calibration strategies. The formation-redshift approach of \citet{Bullock_2001} identifies $z_{\rm coll}$ as the epoch at which a fixed fraction $F$ of the present-day halo mass was assembled, with concentration following from $c \propto [H(z_{\rm coll})/H(z)]^{2/3}$ after \citet{Maccio_2008}. A second class consists of power-law or broken-power-law fits in $M$ or peak height $\nu \equiv \delta_c/\sigma(M,z)$, covering \citet{Duffy_2008}, \citet{klypin_2011}, \citet{Bhattacharya_2013}, \citet{Dutton_2014}, \citet{klypin_2016}, \citet{Child_2018}, and the \citealt{Ludlow_2016} analytic fitting formula (\texttt{ludlow16\_fit}). The third class comprises physically motivated universal models in which $c$ depends on $\nu$ and the local power-spectrum slope $n_{\rm eff}$, including \citet{Prada_2012}, \citet{Diemer_2015}, the MAH-based \citet{Ludlow_2016} model, and \citet{Ishiyama_2021}, which applies the \citealt{Diemer_2019} functional form re-calibrated to the Uchuu simulation. Beyond CDM, the module covers WDM (\citealt{Schneider_2012}) and FDM (\citealt{Laroche_2022}; \citealt{Dentler_2022}), each returning $c = c_{\rm CDM} \times \mathcal{F}(M/M_{1/2})$ where $\mathcal{F}$ is a suppression factor computed through the shared \texttt{suppression\_factor} utility. All models converge to CDM above $M_{1/2}$; below it, the \citealt{Laroche_2022} FDM model can fall below $c=1$, indicating formally unconcentrated haloes where $r_s \geq r_{\rm vir}$, while \citealt{Dentler_2022} predicts a shallower decline.

All CDM outputs were validated against \textsc{colossus} \citep{Diemer_2018}. Figure~\ref{fig:conc} shows five representative CDM models at $z=0$, where \textsc{phantom} results (solid lines) are compared against \textsc{colossus} (dashed lines) for \texttt{bullock01}, \texttt{dutton14}, \texttt{klypin16}, \texttt{ludlow16}, and \texttt{ishiyama21}; each model shares the same colour between the two codes, making deviations immediately visible. The fractional residual $\Delta c/c_{\rm Col}$, shown in the lower sub-panel of Figure~\ref{fig:conc}(a), confirms agreement better than $1$--$2\%$ across $10^{8}$--$10^{15}\,h^{-1}M_\odot$, with the colour-matched pairs tracking each other closely throughout. The remaining CDM models are implemented but omitted for clarity; all pass the same threshold. The WDM and FDM suppression curves, shown in Figure~\ref{fig:conc}(b), converge to the \citet{Ishiyama_2021} CDM baseline above the respective half-mode masses and diverge below them, with \citet{Laroche_2022} falling below $c=1$ at low mass while \citet{Dentler_2022} predicts a shallower decline over the same range.

\begin{table*}
\centering
\footnotesize
\setlength{\tabcolsep}{12pt}
\caption{Implemented density profiles and concentration–mass relations in \textsc{phantom}. The concentration models are selected by their \texttt{key} in the dispatcher, with each row listing the scenario and primary reference. Density profiles are configured directly rather than through the dispatcher, but use the same \texttt{key} naming convention. Full parameter descriptions and usage examples are provided in the \phantomwiki .}
\label{tab:phantom_models_dc}
\small
\resizebox{\textwidth}{!}{%
\begin{tabular}{p{2.4cm} p{1.6cm} p{14.5cm}}
\hline\hline
Model key & Scenario & Notes (Reference) \\
\hline
\multicolumn{3}{l}{\textit{Density profiles}} \\
\texttt{nfw}                  & CDM & Cuspy $r^{-1}(1+r/r_s)^{-2}$ profile; $\rho_s$ from mass conservation within $R_\Delta$ \citep{Navarro_1997}.\\
\texttt{hernquist}            & CDM & Steeper $r^{-1}(1+r/r_s)^{-3}$ outer fall-off; finite total mass $M_\mathrm{tot}=2\pi\rho_s r_s^3$; approximates de Vaucouleurs law \citep{Hernquist_1990}.\\
\texttt{einasto}              & CDM & Exponential profile with shape parameter $\alpha_e=0.155+0.0095\,\nu_\mathrm{vir}^2$, capped at $0.3$; $\nu_\mathrm{vir}$ evaluated at $M_\mathrm{vir}$ via NFW mass-definition conversion \citep{Einasto_1965,Gao_2008,Benson_2011}.\\
\texttt{dk14}                 & CDM & Einasto inner profile with splashback truncation $f_\mathrm{trans}$ and power-law outer envelope $\rho_\mathrm{out}$; $r_t$ from $\nu_{200m}$ at $M_{200m}$ obtained via NFW spherical overdensity conversion \citep{Diemer_2014}.\\
\texttt{soliton}              & FDM & Ground-state Schr\"{o}dinger--Poisson core; $\rho_c$ and $r_c$ from core-halo mass relation; \texttt{cosmo.m22} \citep{Schive_2014_cosmic}.\\
\texttt{composite\_analytic}  & FDM & Soliton for $r\leq r_\times$, NFW for $r>r_\times$; $r_\times$ from analytic intersection; parameters set by core-halo mass relation and concentration model \citep{Robles_2018}.\\
\texttt{composite\_numerical} & FDM & Same composite form fitted directly to a spherically averaged simulation profile; $r_\times$ selected by goodness-of-fit optimisation.\\
\hline
\multicolumn{3}{l}{\textit{Concentration--mass relation}} \\
\texttt{bullock01}      & CDM & Formation-redshift model; \citet{Maccio_2008} revision default \citep{Bullock_2001}.\\
\texttt{duffy08}        & CDM & Power-law in $M$; NFW and Einasto; $M_{\rm 200c/vir/200m}$; WMAP5 \citep{Duffy_2008}.\\
\texttt{klypin11}       & CDM & Double power-law with low-mass upturn; distinct and subhalo samples \citep{klypin_2011}.\\
\texttt{bhattacharya13} & CDM & Power-law in $c$--$\nu$ space; $M_{\rm 200c/vir/200m}$; WMAP7 \citep{Bhattacharya_2013}.\\
\texttt{prada12}        & CDM & $\sigma$-based model with rescaling; captures high-mass upturn \citep{Prada_2012}.\\
\texttt{dutton14}       & CDM & Power-law in $\log M$; $M_{\rm 200c}$ and $M_{\rm vir}$; \textit{Planck}~2013 \citep{Dutton_2014}.\\
\texttt{diemer15}       & CDM & Universal double power-law in $\nu$; depends on $n_{\rm eff}$; any cosmology \citep{Diemer_2015}.\\
\texttt{ludlow16}       & CDM & MAH-based universal model; any mass, redshift, cosmology \citep{Ludlow_2016}.\\
\texttt{ludlow16\_fit}  & CDM & Analytic broken power-law in $\nu$; faster alternative to \texttt{ludlow16} \citep{Ludlow_2016}.\\
\texttt{klypin16}       & CDM & Power-law; $M_{\rm 200c}$ and $M_{\rm vir}$; mass- or $\nu$-based evaluation \citep{klypin_2011}.\\
\texttt{child18}        & CDM & Transition-mass model; individual/relaxed/stacked samples; $z\leq4$ \citep{Child_2018}.\\
\texttt{diemer19}      & CDM & Universal double power-law in $\nu$ and $n_{\rm eff}$; predecessor to \texttt{ishiyama21}; any cosmology \citep{Diemer_2019}.\\
\texttt{ishiyama21}     & CDM & \citet{Diemer_2019} form re-calibrated to Uchuu; CDM reference baseline \citep{Ishiyama_2021}.\\
\texttt{schneider12} & WDM & Power-law suppression below $M_{1/2}$; \texttt{cosmo.m\_wdm\_keV} \citep{Schneider_2012}.\\
\texttt{laroche22}   & FDM & $\mathcal{F}=1+a(M/M_{1/2})^b)^c$; applied to any CDM baseline; \texttt{cosmo.m22} \citep{Laroche_2022}.\\
\texttt{dentler22}   & FDM & Two-factor suppression from \citealt{Kawai_2024}; independent FDM calibration \citep{Dentler_2022}.\\

\hline
\hline
\end{tabular}}
\end{table*}


\section{Examples}
\label{sec:examples}

Sections~\ref{sec:cosmology}--\ref{sec:halo_obs} illustrate each module through validation figures; the two examples below demonstrate use cases that combine multiple layers --- lensing convergence and rotation curves from the observables layer --- to show how a typical analysis workflow proceeds from cosmology initialisation to a science output. Another worked example, based on the large-scale halo bias, is provided in Appendix~\ref{app:halo_bias_example} and mirrors one of the test scripts distributed with the code. The complete script is available in the \textsc{phantom} repository under examples and tests.

\subsection{Gravitational Lensing Convergence}
\label{sec:lensing}

The gravitational lensing convergence $\kappa(R)$ connects the projected dark matter distribution to observable signatures in weak and strong lensing surveys. Given a spherically symmetric density profile $\rho(r)$, the surface mass density at projected radius $R$ is obtained via the Abel transform in Eq.~\ref{eq:Sigma} \citep{Bartelmann_1996, Wright_2000, Binney_2008}. This function is evaluated in profile observable script using the singularity-free substitution $r = \sqrt{R^{2} + t^{2}}$ to remove the integrable divergence at $r = R$ (Section~\ref{sec:halo_obs}). The critical surface density depends only on angular diameter distances to the lens and source \citep{Schneider_1992, Bartelmann_1996, Xin_2013},
\begin{equation}
  \Sigma_{\rm cr} \;=\; \frac{c^{2}}{4\pi G}
  \frac{D_{\rm s}}{D_{\rm l}\,D_{\rm ls}},
  \label{eq:sigma_cr}
\end{equation}
where $D_{\rm l}$, $D_{\rm s}$, and $D_{\rm ls}$ are the angular diameter distances to the lens, to the source, and from the lens to the source, all returned by the cosmology module (Section~\ref{sec:cosmology}). The convergence then follows as,
\begin{equation}
    \kappa(R) = \Sigma(R)/\Sigma_{\rm cr}
    \label{eq:grav_converg}
\end{equation}.
which identifies image deformation due to weak lensing. Eqs.~\eqref{eq:Sigma} and~\eqref{eq:sigma_cr} are identical for every profile model; only the density array $\rho(r)$ changes between calls to the density profile functions.

\begin{figure*}
  \centering
  \includegraphics[width=0.49\textwidth]{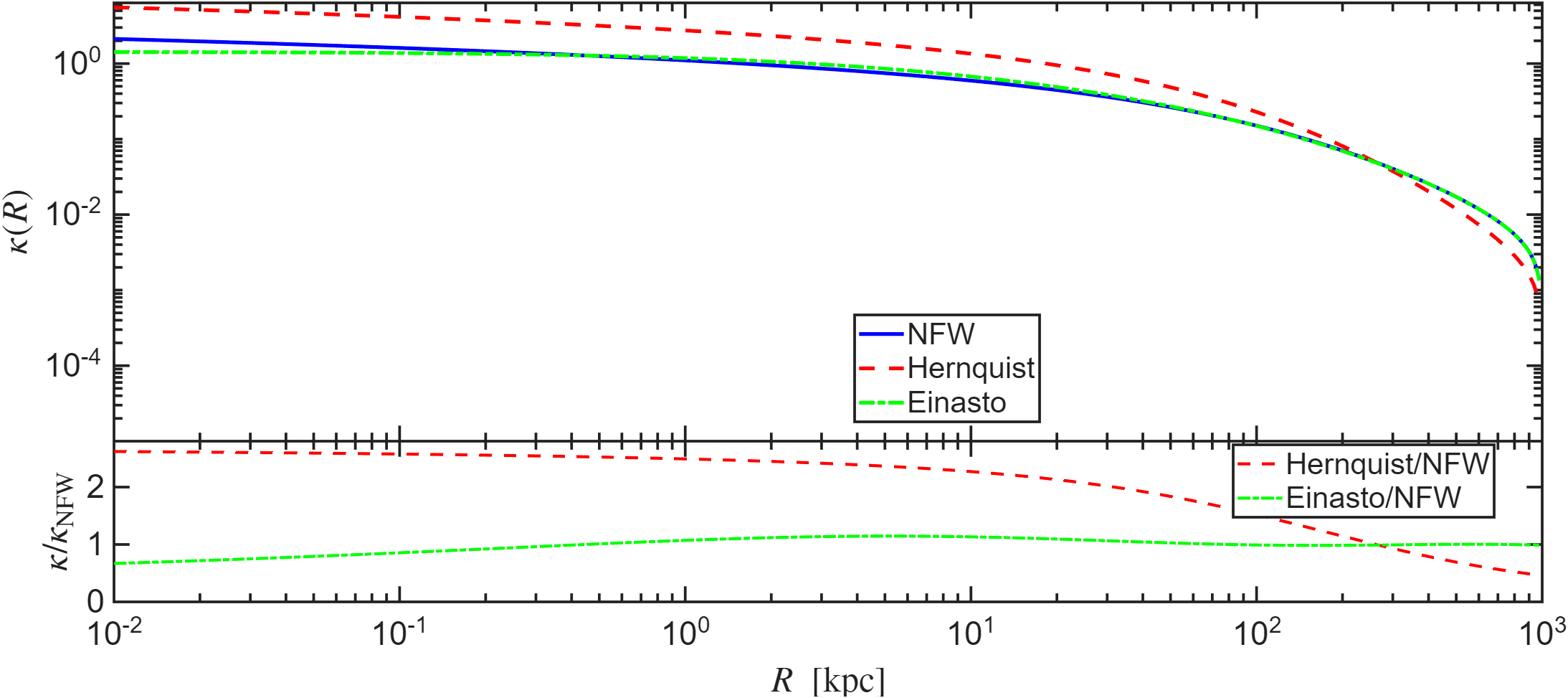}
  \includegraphics[width=0.49\textwidth]{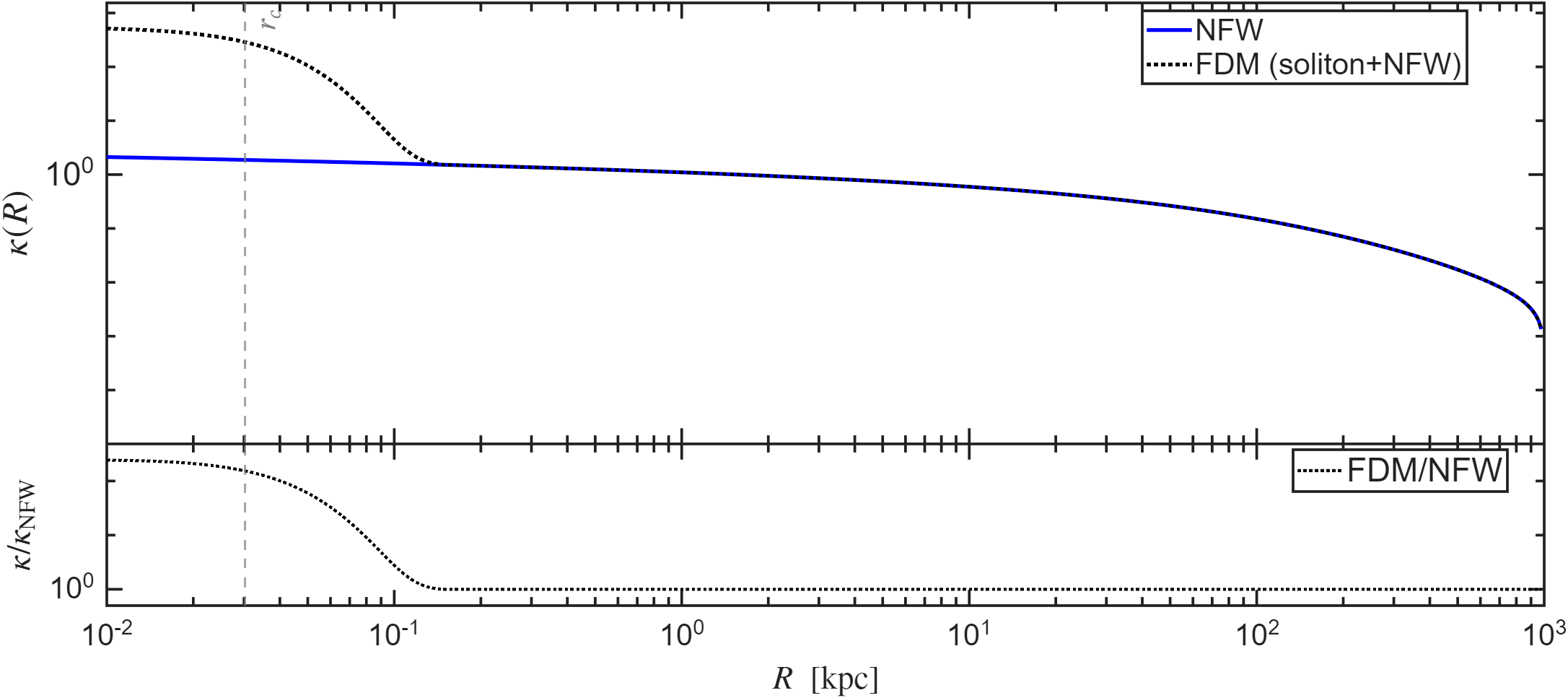}
  \caption{
    Gravitational lensing convergence $\kappa(R) = \Sigma(R)/\Sigma_{\rm cr}$ computed with \textsc{phantom} for a halo of $M_{200\mathrm{c}} = 10^{14}\,h^{-1}\,M_{\odot}$, $c = 5$, $z_{\rm l} = 0.3$, $z_{\rm s} = 1.0$, and Planck18 cosmology. The surface mass density $\Sigma(R)$ is obtained from Eq.~\eqref{eq:Sigma} via profile observable function; $\Sigma_{\rm cr}$ follows from Eq.~\eqref{eq:sigma_cr}. \textit{Left:} CDM profile comparison. The upper sub-panel shows $\kappa(R)$ for the NFW \citep[blue solid;][]{Navarro_1997}, Hernquist \citep[red dashed;][]{Hernquist_1990}, and Einasto \citep[green dash-dot;][]{Einasto_1965}, profiles. The lower sub-panel shows the ratio $\kappa/\kappa_{\rm NFW}$. \textit{Right:} FDM composite (soliton$+$NFW; black dotted) compared against NFW (blue solid), with boson mass $m_{22} = 1$ and soliton parameters from~\ref{subsec:fdm_obs}.The vertical dashed line marks the soliton core radius $r_c$. At $R \lesssim r_c$, the soliton produces a localised convergence enhancement confined to sub-kiloparsec projected radii; beyond the transition radius $r_{\times}$ the ratio $\kappa_{\rm FDM}/\kappa_{\rm NFW}$ converges to unity, recovering the standard NFW profile at large scales.Such a sub-kiloparsec enhancement is the scale at which strong lensing flux ratio anomalies become sensitive to dark matter substructure \citep{Dalal_2002, Laroche_2022}.
  }
  \label{fig:lensing}
\end{figure*}

A $10^{14}\,h^{-1}\,M_{\odot}$ halo at $z_{\rm l} = 0.3$ with concentration $c = 5$ and source redshift $z_{\rm s} = 1.0$ (Planck18 cosmology) is used throughout. The left panel of Figure~\ref{fig:lensing} shows $\kappa(R)$ for three CDM profiles implemented in \textsc{phantom}. NFW and Einasto track each other across most of the radial range, with the Einasto profile lying marginally above NFW at small $R$ owing to its cusp-free exponential interior \citep{Einasto_1965, Montenegro_2012}. The Hernquist profile \citep{Hernquist_1990} exceeds NFW at all projected radii because its steeper outer fall-off ($\rho \propto r^{-4}$) concentrates more mass along any line of sight; the ratio $\kappa_{\rm Hernquist}/\kappa_{\rm NFW}$ reaches $\sim$2.6 near $R = 0.01\,\mathrm{kpc}$ and falls below unity only beyond $R \approx 270\,\mathrm{kpc}$ as the steeper outer envelope reduces the projected mass relative to NFW. All three profiles converge at intermediate radii where their shapes are nearly identical.

The right panel of Figure~\ref{fig:lensing} isolates the effect of the FDM soliton core by comparing the composite (soliton$+$NFW) profile against the pure NFW for the same halo parameters. The boson mass is $m_{22} \equiv m/(10^{-22}\, \mathrm{eV}) = 1$, and the soliton parameters are set through the core-halo mass relation of \citet{Schieve_2014_core} (Section~\ref{subsec:fdm_obs}). The vertical dashed line marks the resulting soliton core radius $r_c \approx 0.15\,\mathrm{kpc}$. At $R \ll r_c$, the soliton's near-constant-density interior produces a localised excess in $\kappa$ relative to the pure NFW profile, confined to projected radii $R \lesssim r_c \approx 0.15\,\mathrm{kpc}$; this enhancement would manifest as perturbations to flux ratios in quadruple-image strong lens systems \citep{Laroche_2022, Oguri_2026}. Beyond the soliton-to-NFW transition radius $r_{\times}$, the composite profile is indistinguishable from a pure NFW halo, and the ratio $\kappa_{\rm FDM}/\kappa_{\rm NFW}$ in the lower sub-panel converges to unity at $R \gtrsim 1\,\mathrm{kpc}$. The convergence profiles computed here with \textsc{phantom} provide the mass-projection ingredient required for such tests. Strong gravitational lensing, through flux ratio anomalies in quadruple-image systems, offers one of the few direct probes of dark matter structure on sub-galactic scales \citep{Dalal_2002,Laroche_2022}, and the soliton-driven enhancement in $\kappa$ at $R \lesssim r_c$ is in principle a distinguishing signature between FDM and CDM halo models — provided sufficiently high-resolution lensing data are available.

\subsection{Circular velocities for SPARC galaxies}
\label{sec:Vc_examples}

As another example, we construct circular-velocity curves for two well-studied SPARC galaxies, the late-type spiral NGC~2403 and the dwarf UGCA~442 \citep{Lelli_2016}.  This workflow exercises the halo-structure and observables layers by combining an NFW or soliton and NFW density profile with the concentration model of \citet{Ishiyama_2021}, using the \texttt{ishiyama21} calibration in the \textsc{phantom} concentration dispatcher and computing $V_{\rm c}(r)$ through the enclosed-mass routine of Section~\ref{sec:halo_obs}.

\begin{figure*}
    \centering
    \includegraphics[width=0.48\textwidth]{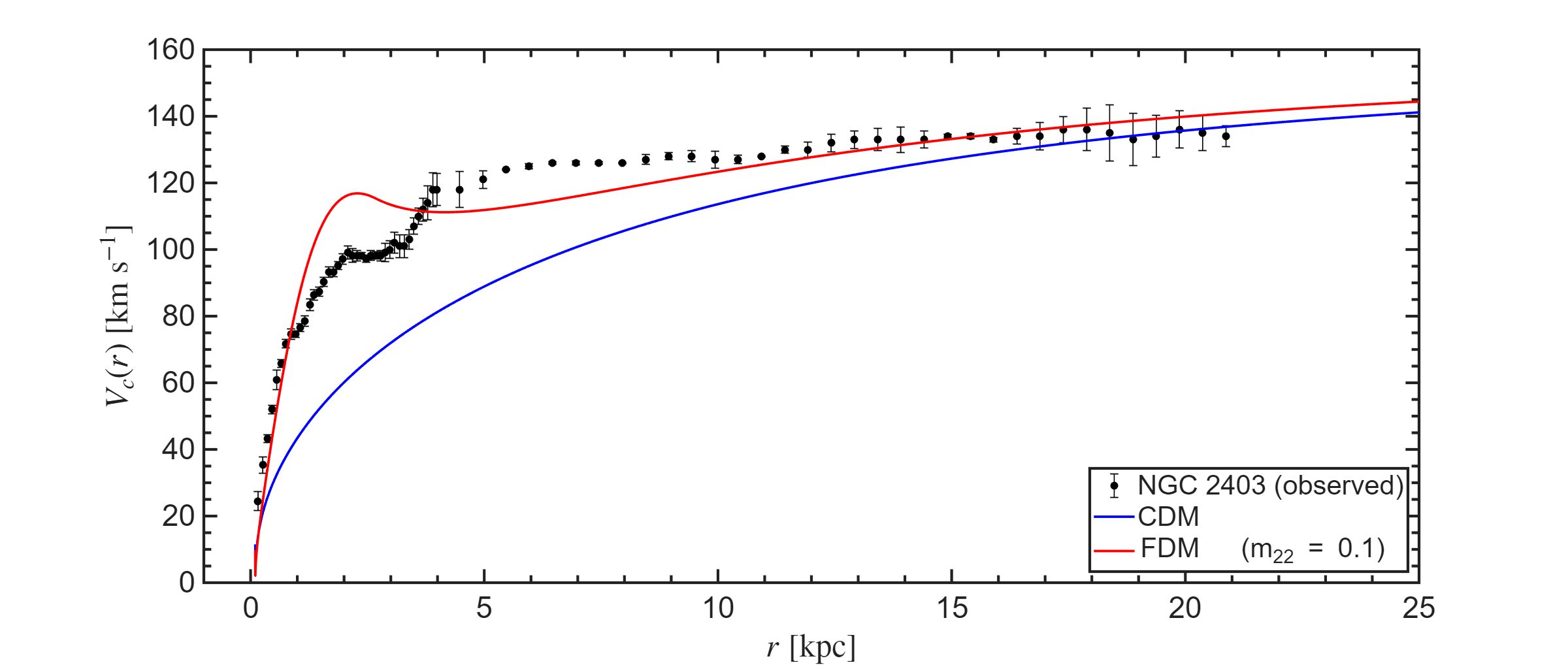}%
    \hfill
    \includegraphics[width=0.48\textwidth]{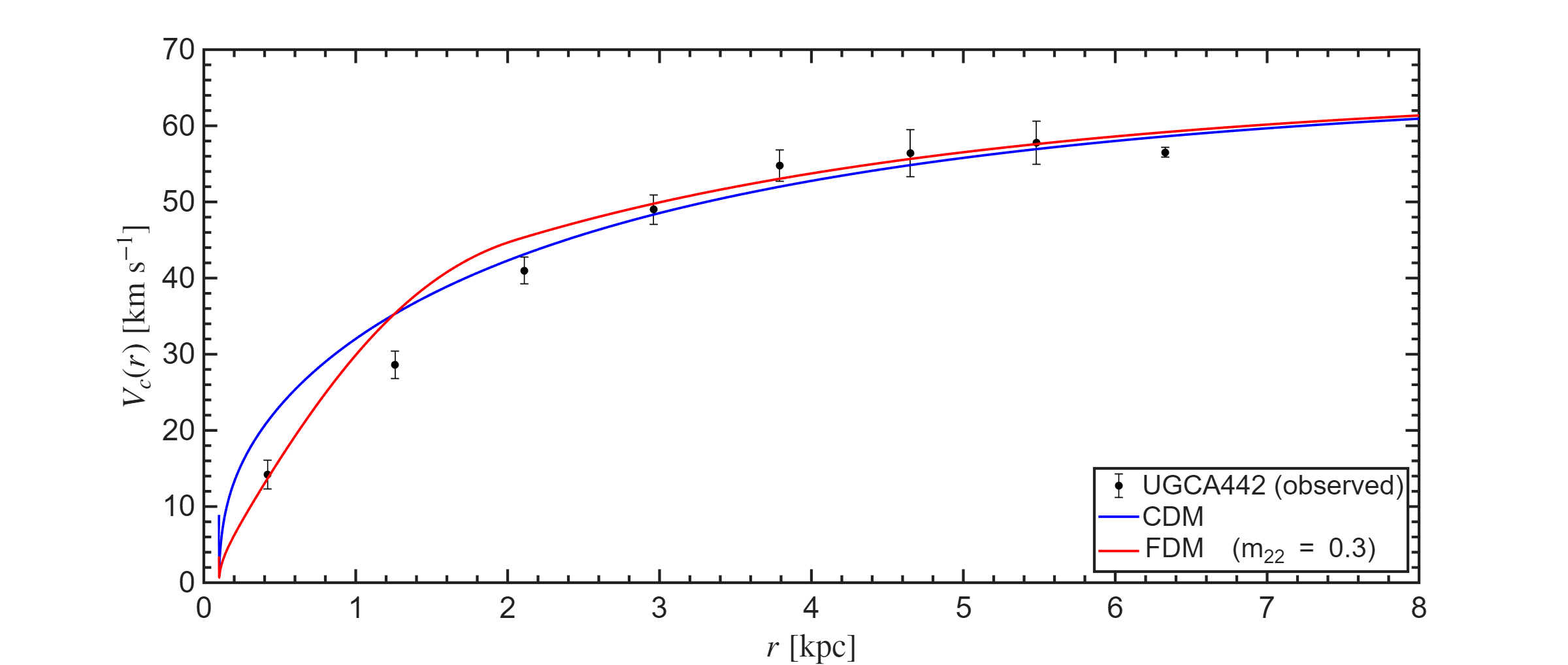}
    \caption{Circular-velocity curves computed with \textsc{phantom} compared to SPARC rotation-curve data \citep{Lelli_2016}.
    \textit{Left:} NGC~2403, a late-type spiral with halo mass
    $M_{200} = 9.4\times10^{11}\,M_\odot$.
    \textit{Right:} UGCA~442, a late-type dwarf with halo mass
    $M_{200} = 6\times10^{10}\,M_\odot$.
    In each panel the blue line shows a pure NFW profile, and the red line shows the soliton+NFW composite for FDM with boson mass $m_{22}=0.1$ (left) and $m_{22}=0.3$ (right), using the core--halo relation of \citet{Schieve_2014_core}. Concentrations are from \citet{Ishiyama_2021} at $z=0$. The central velocity excess in the FDM curves traces the solitonic core; the larger core in NGC~2403 follows directly from the lower boson mass. The halo masses and boson masses are parameter choices made for visual agreement with the data and are not constraints on the dark matter content of these systems. The complete script that generates this figure is provided in the \textsc{phantom} repository under \texttt{examples/}.}
    \label{fig:Vc_sparc}
\end{figure*}

For NGC~2403, we adopt $M_{vir} = 9.4\times10^{11}\,M_\odot$ and for UGCA~442, we adopt $M_{vir} = 6\times10^{10}\,M_\odot$; concentrations are drawn from the \citet{Ishiyama_2021} CDM relation at $z=0$ using the virial overdensity definition. In each panel of Figure~\ref{fig:Vc_sparc} the blue curve shows the pure NFW circular velocity and the red curve shows the soliton and NFW composite with a boson mass $m_{22}=0.1$ (NGC~2403) and $m_{22}=0.3$ (UGCA~442) (i.e. $m_\psi = m_{22}\times10^{-22}\,{\rm eV}$), with the core--halo mass relation of \citet{Schieve_2014_core}. Both galaxies are reproduced at the outer radii by the CDM model, while the FDM composite produces a distinct central bump that reflects the solitonic core; the bump is more pronounced in NGC~2403 because the lower boson mass yields a larger core radius. In both cases the circular velocities are obtained by a single call to \texttt{halo\_obs} with the chosen mass, overdensity, and profile key, demonstrating how the cosmology, concentration, and profile layers combine in a typical \textsc{phantom} workflow.

These examples are deliberately simple and omit a full baryonic mass model or a parameter exploration of the halo mass--concentration degeneracy.  The halo masses and FDM boson masses quoted here should therefore be viewed as illustrative parameter choices that demonstrate how \textsc{phantom} connects halo profiles to observed rotation curves, rather than as measurements of the dark-matter content of UGCA~442 or NGC~2403.

\section{Summary and future development}
\label{sec:conclusion}

\textsc{phantom} provides a validated \textsc{matlab} toolbox, with a parallel \textsc{octave} distribution, for dark matter halo analysis, from linear field statistics to halo observables for CDM, WDM, and FDM cosmologies. All core routines agree with \textsc{colossus}, \textsc{hmf}, and \textsc{halomod} at the sub-percent level for shared models, and the FDM soliton–halo composite profiles extend the available functionality beyond the Python ecosystem. The code is publicly available under the MIT licence at the GitHub
repository.

\textsc{phantom} fills a gap in the current software ecosystem. Validated toolkits such as \textsc{colossus} \citep{Diemer_2018}, \textsc{hmf} \citep{Murray_2013}, \textsc{halomod} \citep{Murray_2021}, and \textsc{cluster-toolkit} \citep{McClintock_2019} are Python based, while many
observational and instrumentation workflows in gravitational lensing, telescope operations, and spectroscopic fitting remain in \textsc{matlab}. Porting those pipelines to Python carries a non-trivial validation cost. By implementing the same halo physics in a native \textsc{matlab} environment, \textsc{phantom} lets users call validated cosmology, halo statistics, and halo-structure models inside existing \textsc{matlab} code. The dispatcher architecture, in which all modules accept a string key and a cosmology structure, allows models to be swapped or extended without changing calling scripts.

Several extensions are planned or open to contributions. On the physics side, a halo-model module that combines the existing bias and profile machinery into a non-linear matter power spectrum is a natural next step. Mass function and concentration models for self-interacting dark matter (SIDM) and mixed dark matter, non-spherical and triaxial profiles for cluster lensing, and subhalo abundance functions would broaden the scope toward satellite and group-scale applications. On the numerical side, a full FFTLog pipeline for the halo-model power spectrum would complement the current direct-integration correlation function, and a \textsc{class} bridge would extend the present CAMB and axionCAMB interfaces to a wider set of cosmological models.

Community contributions are actively encouraged and follow the workflow documented in the repository. Contributors are expected to use a fork–branch–pull-request model to add
unit tests in the \texttt{tests/} directory for every new function, and to match the units and calling conventions used elsewhere in the code. New models should be implemented as dispatcher-compatible functions that accept the relevant input (typically $\sigma(M,z)$ and a cosmology structure) and return the quantity of interest, and they should be accompanied by a validation figure or comparison against an independent code or published tables. Bug reports, feature requests, and proposals for new physics or numerical options are handled through the GitHub issue tracker and are reviewed with the same emphasis on numerical accuracy and clear physical motivation.

\section*{Acknowledgments}
The author thanks the developers of colossus~\citep{Diemer_2018} for making their Python implementation publicly available, which served as the primary validation benchmark throughout this work. The author also acknowledges the developers of CAMB (Lewis et al. 2000) and axionCAMB (Hlozek et al. 2015), whose outputs are used as optional power-spectrum backends in phantom. This work received no external funding and was carried out as an independent project.


\appendix

\section{Growth factor: auxiliary solvers}
\label{app:growth}

The other three growth factor solvers used by \textsc{phantom} beyond the default Heath-Peebles integral (Eq.~\ref{eq:heath}) is explained in this section.For flat $\Lambda$CDM with negligible radiation, Eq.~\ref{eq:growth_ode} admits the computationally inexpensive approximation of \citet{Eisenstein_1999}. Writing $E^2(a) \equiv \Omega_m a^{-3}$ and defining the time-dependent fractions
\begin{equation}
\Omega_m(a) = \frac{\Omega_m a^{-3}}{E^2(a)},
\qquad
\Omega_\Lambda(a) = \frac{\Omega_\Lambda}{E^2(a)},
\label{eq:omega_a}
\end{equation}
the growth suppression relative to an Einstein-de Sitter universe is captured by a fitting function $g(a)$ depending only on $\Omega_m(a)$ and $\Omega_\Lambda(a)$, giving
\begin{equation}
D(a) = a\,\frac{g(a)}{g(1)}.
\label{eq:growth_eh}
\end{equation}
This path is the default for standard flat $\Lambda$CDM because it avoids numerical integration entirely.

When radiation is non-negligible, the matter-only growth equation breaks down at high redshift. The code therefore combines the Heath-Peebles integral (Eq.~\ref{eq:heath}) at $z \leq 5$ with the analytic approximation of \citet{Gnedin_2018} in the matter-radiation regime,
\begin{equation}
\begin{split}
    D_{\rm G}(a) = a + \frac{2}{3}a_{\rm eq} &+ \frac{a_{\rm eq}}{2\ln 2
    - 3} \\ &\times \left[2\sqrt{1+x} 
    + \left(\frac{2}{3}+x\right)
      \ln\frac{\sqrt{1+x}-1}{\sqrt{1+x}+1}\right],
    \label{eq:gnedin}
\end{split}
\end{equation}
and interpolates between the two solutions in $\ln a$ over $5 \leq z \leq 20$, enforcing a smooth, monotonic growth history from radiation domination through the late $\Lambda$-dominated phase.

For dark energy with a constant or evolving equation of state $w(a) = w_0 + w_a(1-a)$ (CPL parametrisation, \citealt{Chevallier_2001, Linder_2003}), the growth problem is recast via $G(a) \equiv D(a)/a$ following \citet{Linder_2003},
\begin{equation}
\frac{d^2G}{da^2} + A(a)\frac{dG}{da} + B(a)\,G = 0,
\label{eq:growth_cpl}
\end{equation}
where, defining the matter-to-dark-energy ratio $X(a)=(\Omega_{\rm m}/a^3)/(\delta H^2/H_0^2)$ with $\delta H^2/H_0^2\equiv E^2(a)-\Omega_{\rm m}/a^3$,
\begin{equation}
\begin{split}
    A(a) &= \frac{3.5-1.5\,w(a)/(1+X(a))}{a}\,,\\
    B(a) &= \frac{1.5\,[1-w(a)]}{(1+X(a))\,a^2}\,.
\end{split}
\label{eq:growth_cpl_coeff}
\end{equation}
Eq.~\ref{eq:growth_cpl} is integrated numerically with initial conditions $G(a_{\rm min}) = 1$, $dG/da = 0$ deep in the matter-dominated era, and the growth factor is recovered as $D(a) = a\,G(a)$, normalised at $a = 1$.

\section{Variance: filter options}
\label{app:variance}

This appendix collects the smoothing filters available through \texttt{cosmo.varianceFilter}. In all non-top-hat cases, a mass-calibration parameter $c$ centres only the $M$--$R$ relation, $M = \frac{4}{3}\pi\bar{\rho}_{m,0}(cR)^3$, not the window function itself.

The Gaussian filter is
\begin{equation}
W_{\rm G}(k,R) = \exp\!\left(-\frac{k^2R^2}{2}\right),
\label{eq:gaussian}
\end{equation}
and is provided for cases where smooth behaviour or derivatives of $\sigma(R)$ are needed. It has no compact real-space kernel but produces well-behaved derivatives of the variance.

The sharp-$k$ filter is
\begin{equation}
W_{\rm sk}(k,R) = \Theta(1 - kR),
\label{eq:sharpk}
\end{equation}
where $\Theta$ is the Heaviside step function. This filter yields Markovian random walks in the variance as a function of scale \citep{Bond_1991,Diemer_2018} and is included for excursion-set applications. It has no compact real-space kernel and its mass-radius mapping is ambiguous, so it is not recommended for general halo statistics. For models with gradual small-scale power suppression, such as WDM, the sharp-$k$ boundary introduces artefacts in mass-function calculations.

The smooth-$k$ filter of \citet{Leo_2018} replaces the Heaviside step with a Lorentzian-like suppression,
\begin{equation}
W_{\rm smk}(k,R) = \frac{1}{1 + (kR)^{2\beta}},
\qquad \beta = 4,
\label{eq:smoothk}
\end{equation}
which transitions smoothly from unity to zero around $kR = 1$. This avoids the sharp-$k$ artefacts in WDM mass-function calculations and is the recommended choice for models with a smooth small-scale cutoff.

For models with more complex cutoffs, such as dark acoustic oscillations or FDM-inspired suppressions, the variable-slope smooth-$k$ filter of \citet{Rocamora_2026} allows the slope to
evolve with scale,
\begin{equation}
W_{\rm vsmk}(k,R) =
  \frac{1}{1 + (kR)^{f(kR)}},
\qquad
f(kR) = 2\,\frac{1 + \alpha_1(kR)^{\gamma}}{1 + \alpha_2(kR)^{\gamma}},
\label{eq:vsmk}
\end{equation}
where $\alpha_1 = 14.8$, $\alpha_2 = 3.6$, and $\gamma = 2.1$ are calibrated from simulations. This filter is recommended for models with scale-dependent damping, including WDM and FDM.


\section{Composite Profile Fitting Procedure}
\label{app:fitting}

The composite FDM density profile is constructed by fitting the soliton core and NFW outer halo to distinct radial regions of the simulation data, then stitching the two components at their intersection. The full procedure is implemented in the \textsc{phantom} functions
\texttt{soliton\_nfw\_composite}, \texttt{find\_best\_m}, and
\texttt{fit\_profile\_generic}. The soliton core follows Eq.~\eqref{eq:soliton_profile}) and the outer halo  follows the NFW form of Eq.~\eqref{eq:NFW}), each with two free  parameters ($\rho_0$, $r_c$ and $\rho_s$, $r_s$ respectively),  giving four parameters in total for the composite profile.

Neither profile is valid across the full radial range: the soliton dominates only the inner core and the NFW form describes only the outer halo. The boundary between these regions is set by the cutoff radius $r_\mathrm{cut} = m\,r_c$, where $r_c$ is an initial estimate of the core radius supplied by the user and $m$ is a dimensionless multiplier. The soliton is fitted to data with $r \leq r_\mathrm{cut,sol}$, and the NFW profile is fitted to data with $r_\mathrm{cut,nfw} < r \leq R_\mathrm{vir}$.


The optimal multiplier $m$ is determined by a one-dimensional scan over $m \in [m_\mathrm{min},\,m_\mathrm{max}]$ using 100 uniformly spaced values. At each trial value, \texttt{fit\_profile\_generic} performs a least-squares fit in $\log_{10}$-density space and evaluates a goodness-of-fit statistic; the user may select among reduced chi-squared $\chi^2_\nu = \nu^{-1}\sum[\log_{10}\rho_\mathrm{model} - \log_{10}\rho_\mathrm{data}]^2$, RMSE, AIC, or BIC, with $\chi^2_\nu$ as the default. The value of $m$ minimising the chosen statistic is adopted as the optimal cutoff; using $\chi^2_\nu$ by default prevents the scan from favouring large cutoffs with few data points, since the degrees-of-freedom denominator $\nu = N - k$ penalises underdetermined fits. The optimisation itself is performed in linear density space using \textsc{Matlab}'s \texttt{lsqcurvefit}.

After the best-fit parameters $(\rho_0, r_c)$ and $(\rho_s, r_s)$ are obtained, the transition radius $r_\times$ is located as the outermost crossing of the two profiles on a dense logarithmic grid of 5000 points spanning $[r_\mathrm{min},\,R_\mathrm{vir}]$. The crossing is found using \texttt{polyxpoly} (MATLAB Mapping Toolbox) when available; if the toolbox is absent, \textsc{phantom} falls back to sign-change detection followed by \texttt{fzero} for sub-bin precision. Under Octave, the equivalent \texttt{intersectPolylines} function is used in place of \texttt{polyxpoly}. The composite profile is then
\begin{equation}
  \rho(r) =
  \begin{cases}
    \rho_\mathrm{sol}(r) & r \leq r_\times, \\
    \rho_\mathrm{NFW}(r) & r > r_\times,
  \end{cases}
  \label{eq:composite_app}
\end{equation}
which is continuous at $r_\times$ by construction, since both components return the same density there. Default scan ranges are $m \in [2.0,\,3.5]$ for the soliton and $m \in [3.5,\,10.0]$ for the NFW component; if no intersection is found, the code can optionally extend the scan range iteratively up to a user-specified number of retries.

\section{Additional example: halo bias}
\label{app:halo_bias_example}

The large-scale halo bias $b(M,z)$ quantifies the excess clustering of haloes relative to the underlying matter field and is a key ingredient in halo-model predictions of galaxy clustering, galaxy--galaxy lensing, and the nonlinear matter power spectrum \citep[e.g.][]{Cooray_2004, Tinker_2010}. Given the variance of the linear density field $\sigma(M,z)$ and the collapse threshold $\delta_{\rm c}$, analytic models express the bias as a function of peak height $\nu \equiv \delta_{\rm c}/\sigma$ \citep{Cole_1989, Sheth_1999, Sheth_2001}. The halo-bias module in \textsc{phantom} implements several such fits and exposes them through a unified dispatcher (Section~\ref{sec:hmf_bias}).

For a fixed cosmology, $\sigma(M,z)$ is obtained from the cosmology module (Section~\ref{sec:cosmology}) via the linear power spectrum and real-space top-hat filter (Section~\ref{sec:variance}). The user then selects a bias prescription via the string key \texttt{model} in the call
\begin{equation}
  b(M,z) \;=\; \texttt{halo\_bias}(\texttt{model},\,\sigma(M,z),\,\delta_{\rm c},\,\ldots),
\end{equation}
where additional arguments (e.g.\ overdensity $\Delta$, redshift $z$, cosmology) are passed transparently to the underlying implementation. This interface removes the need to track model-dependent parameterisations of $\nu$ and ensures a consistent treatment of $\delta_{\rm c}$ and $\sigma(M,z)$ across all bias fits.

\begin{figure*}
  \centering
  \includegraphics[width=0.49\textwidth]{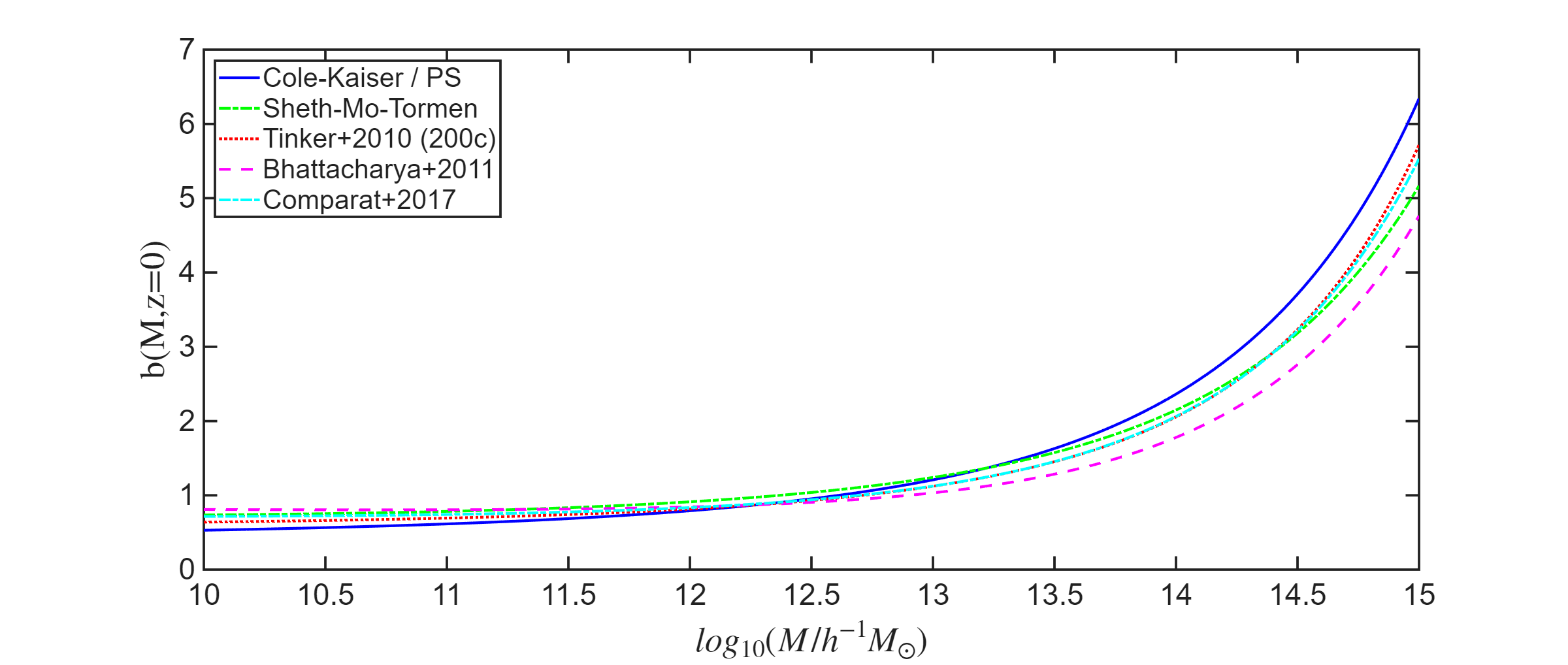}
  \includegraphics[width=0.49\textwidth]{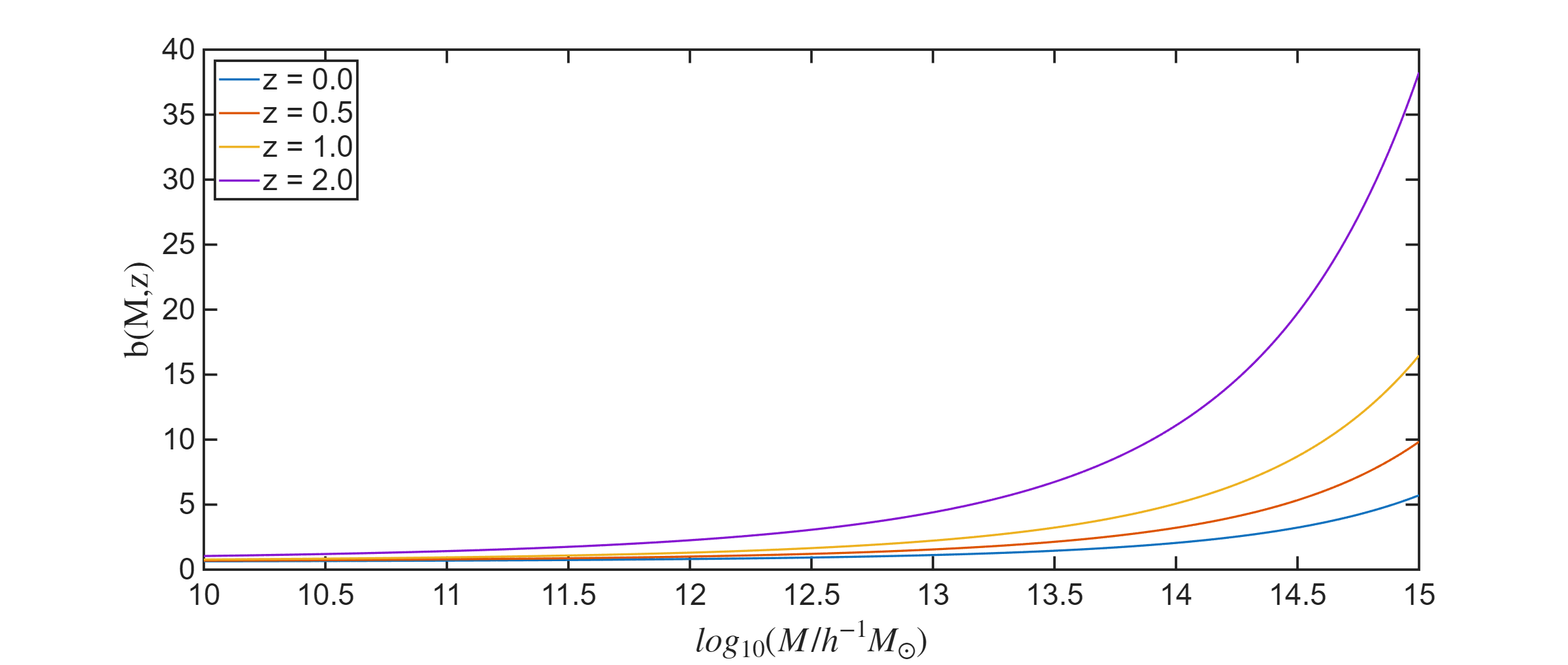}
  \caption{
    Linear halo bias $b(M,z)$ computed with \textsc{phantom} using the unified halo-bias interface.
    \textit{Left:} $b(M)$ at $z=0$ for a Planck18 cosmology, comparing the Press--Schechter/Cole--Kaiser \citep[blue solid;][]{Cole_1989}, Sheth \& Tormen \citep[orange dashed;][]{Sheth_1999}, Sheth--Mo--Tormen moving-barrier \citep[green dash-dotted;][]{Sheth_2001}, and Tinker et al.\ \citep[red dotted;][]{Tinker_2010} prescriptions.
    \textit{Right:} Redshift evolution of the Tinker et al.\ bias at $z = 0$ (blue), $0.5$ (orange), $1$ (green), and $2$ (red). At fixed mass, the bias increases towards higher redshift as haloes of a given mass correspond to rarer peaks in the linear density field.
  }
  \label{fig:bias_example}
\end{figure*}

Figure~\ref{fig:bias_example} illustrates $b(M)$ for a Planck18 cosmology at $z=0$ over the mass range $10^{10}$--$10^{15}\,h^{-1}M_\odot$. The left panel compares four commonly used prescriptions: the Press--Schechter/Cole--Kaiser model \citep[blue solid;][]{Cole_1989}, the Sheth \& Tormen ellipsoidal-collapse fit \citep[orange dashed;][]{Sheth_1999}, the moving-barrier extension of \citet[green dash-dotted]{Sheth_2001}, and the Tinker et al.\ \citeyearpar{Tinker_2010} calibration (red dotted), which is the default choice in \textsc{phantom}. All models predict $b \simeq 1$ near the characteristic mass scale, but diverge systematically at the low- and high-mass ends; this spread is relevant when interpreting highly biased tracers such as massive clusters or very luminous galaxies.

The right panel shows the redshift evolution of the Tinker et al.\ bias for $z = 0,\,0.5,\,1,$ and $2$. At fixed mass, the bias increases monotonically with redshift as haloes of a given mass correspond to rarer peaks in the linear density field. This example demonstrates how the cosmology, variance, and halo-bias modules combine to produce $b(M,z)$ with only a few lines of user code. The full script is provided in the \textsc{phantom} repository under examples.

\bibliographystyle{elsarticle-harv}   
\bibliography{references}             


\end{document}